\documentclass{WileyASNA-v1}

\usepackage{graphicx}
\usepackage{amsmath}
\usepackage{color}
\usepackage{float}
\usepackage{lscape}
\usepackage{longtable}
\usepackage{rotating}
\usepackage{ulem}
\usepackage{tabularx}
\usepackage{hyperref}
\newcolumntype{Y}{>{\centering\arraybackslash}X}

\articletype{Research Article}%

\received{03.02.2025}
\revised{31.10.2025}
\accepted{05.11.2025}

\begin{document}

\title{Investigating Circumstellar Atomic Radiation-driven Dynamics}

\abstract{The interactions between stars and their orbiting planets, driven by forces such as stellar radiation and gravity, play an essential role in shaping exoplanetary atmospheres and gas-rich debris discs. One way to look into the composition of these environments is to observe how they can contaminate the stellar photospheres. For that, we examine how stellar radiation pressure and gravity influence atomic species and analyse their effects across various stellar effective temperatures. Using the radiative-to-gravitational force ratio, we determined the atomic movement direction and assessed the velocity boost imparted to neutral atoms escaping from exoplanet atmospheres or debris discs. Incorporating the solar far ultraviolet/extreme ultraviolet spectrum to address flux discrepancies in the {\sc{atlas9}} model, we find that radiation affects atoms differently according to their ionisation state, with highly ionised species less affected by stellar radiation. Our results conclude that the stars most suitable for observing stellar contamination are those between 6,500 and 8,000 K, with neutral noble gases and ionised iron-peak elements as the most likely contaminants.}

\author[1]{Alexandra Lehtmets*}

\author[2,1]{Mihkel Kama}

\author[3]{Luca Fossati}

\author[1]{Anna Aret}

\authormark{A. Lehtmets \textsc{et al}}

\address[1]{\orgdiv{Tartu Observatory}, \orgname{University of Tartu}, \orgaddress{\state{Tõravere}, \country{Estonia}}}

\address[2]{\orgdiv{Department of Physics and Astronomy}, \orgname{University College London (UCL)}, \orgaddress{\state{London}, \country{United Kingdom}}}

\address[3]{\orgdiv{Space Research Institute}, \orgname{Austrian Academy of Sciences}, \orgaddress{\state{Graz}, \country{Austria}}}

\corres{*Alexandra Lehtmets, University of Tartu, Tartu Observatory, 61602 Tõravere, Estonia. \email{alexandra.lehtmets@ut.ee}}

\keywords{acceleration of particles; (stars:) circumstellar matter; (stars:) planetary systems, scattering}

\jnlcitation{\cname{%
\author{A. Lehtmets}, 
\author{M. Kama}, 
\author{L. Fossati}, and 
\author{A. Aret}} (\cyear{2025}), 
\ctitle{Investigating Circumstellar Atomic Radiation-driven Dynamics}, \cjournal{Astronomische Nachrichten}.}

\maketitle


\begin{figure*}[t]
\centerline{\includegraphics[width=0.98\textwidth]{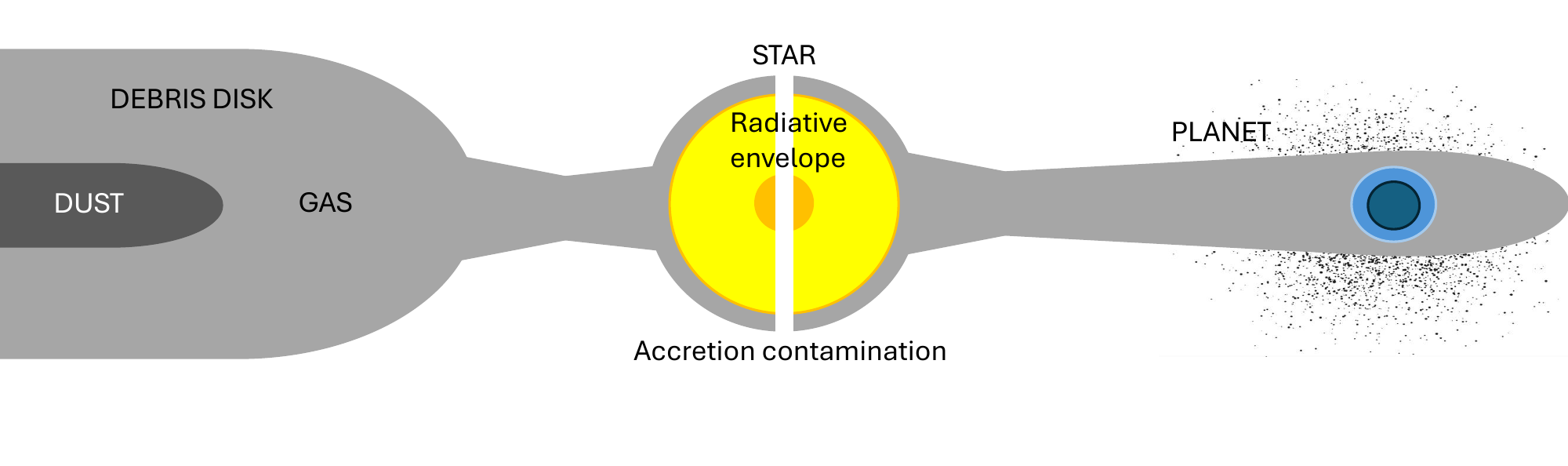}}
\caption{Illustration showing different scenarios of gaseous circumstellar material spreading around a star and either accreting onto or being accelerated away from it. \emph{Left-hand panel:} A gas-rich debris disc (see Section\, \ref{disc}). \emph{Right-hand panel:} An evaporating giant or rocky exoplanet (see Sections\, \ref{jupiter}\,\&\,\ref{rocky}).
\label{fig:graph}}
\end{figure*}

\section{Introduction}\label{intro}

Tenuous circumstellar gas appears in various scenarios, such as planet formation and evolution. Some gas-rich debris discs can retain primordial or secondary gas \citep{nakatani2023, kral}, while strongly irradiated exoplanets can release gas through atmospheric escape \citep{vidal-madjar, haswell,lieshout1, lieshout2}. The evolution of circumstellar gas is particularly significant for early-type stars, with high luminosities that enhance atmospheric escape or icy planetesimal sublimation and reduced stellar envelope convection, making contamination more likely \citep{turcotte, jermyn}.

Accretion contamination becomes prominent in stars with weak mixing processes and low stellar wind activity. The accreted material is rapidly diluted or repelled for stars with vigorous internal mixing or active stellar winds. Consequently, contamination is primarily associated with early-type stars, such as F- and cooler A-type stars. Understanding how gas behaves within discs or evaporates from exoplanet atmospheres is important for identifying potential stellar photospheric contamination.

The peculiar strong depletion of dust-forming elements in the photospheric abundance patterns of $\lambda\,$Bo\"{o} stars \citep{paunzen} has been hypothesised to arise from the accretion of circumstellar material. This material could originate from the interstellar medium \citep{kamp}, a protoplanetary disc \citep{kama}, or potentially from a nearby hot Jupiter with an escaping atmosphere \citep{jura}. Models of stellar mixing, interstellar medium (ISM) accretion, disc accretion, and planet evaporation rates \citep{saffe} point to protoplanetary discs as the most likely source of significant stellar contamination \citep{jermyn}, while observations of early-type stars with hot Jupiters have thus far not found any accretion contamination signatures \citep{saffe}.

\citet{fernandez} investigated circumstellar gas dynamics by analysing the interaction between radiation and gravitational forces on atomic species in gas debris disc around $\beta$ Pictoris. Certain chemical species' transition lines occur in the far ultraviolet/extreme ultraviolet (FUV/EUV) bands. In the case of $\beta$ Pictoris, the EUV radiation originates from chromospheric and coronal emissions, similar to those observed in the modern Sun \citep{Wu2025}.

Our study aims to generalise this approach by calculating the parameter $\beta$ (which is the relationship between stellar gravity and radiation pressure) for the first $28$ atomic species in three ionisation states (neutral, singly ionised and doubly ionised), spanning a wide range of stellar effective temperatures from $4,000$ to $19,000$\,K. We also evaluate the velocity boost experienced by neutral atoms before ionisation and explore the impact of high-energy stellar emission. Radiation pressure coefficients have also been previously reported by \cite{burns1979,beust1993,liseau,aumic, jura}.


In this study, we explore three scenarios in which our model can be applied to understand circumstellar gas (see Figure \ref{fig:graph}):
\begin{enumerate}
    \item Debris discs are observed around stars of various spectral types, including A, F, G, K, and M-type stars \citep{lestrade}. Although the gas mass in these discs is significantly lower than that in protoplanetary discs, some debris discs have been found to harbour notable amounts of gas. This gas can have two possible origins: it may be a residual gas dominated by hydrogen left over from the protoplanetary phase \citep{nakatani2023}, or it may be a secondary gas, often dominated by carbon and oxygen, sublimated from icy bodies within the disc \citep{kral}. Accurate modelling of these gas-rich but tenuous debris discs requires consideration of radiation pressure effects on individual atomic components.
    \item Hot Jupiters, predominantly found around G and K-type stars \citep{johnson}, include some extreme cases like KELT-9\,b, which orbits a host star with an effective temperature of approximately $10,000$\,K \citep{gaudi}. Close-in hot Jupiters often exhibit tails of evaporated gas, observable as either forward- or backwards-facing structures \citep{vidal-madjar, haswell}. These gas tails are driven by intense stellar ultraviolet radiation, leading to atmospheric escape rates ranging from $10^{10}$\,g\,s$^{-1}$ for HD\,209458\,b \citep{Vidal-Madjar2003} to as high as $10^{12}$\,g\,s$^{-1}$ for ultra-hot Jupiter KELT-9\,b \citep{gaudi, borsa2022}. Close-in hot Jupiters generally have orbital periods shorter than two days and are found around relatively young stars. There is an ongoing discussion about "catastrophic evaporation," where extreme atmospheric escape may leave behind only a planetary core or nothing at all \citep{hebrard}.
    \item Close-in rocky planets are commonly detected around K and M-type stars \citep{arabhavi}. Studies such as \citet{lieshout1, lieshout2} reveal that rocky exoplanets can exhibit tails composed of evaporated atmospheric material, including dust. For example, the dust ejected from planets like KIC\,12557548\,b and KOI-2700\,b has been found to contain elements such as aluminium, oxygen, iron, and silicon. The estimated dust-loss rates for these planets range between $1.33 \times 10^9$ and $3.22 \times 10^{11}$\,g\,s$^{-1}$. Furthermore, some fraction of the dust may sublimate into gas, contributing to the circumstellar environment. Observations, primarily from the Kepler mission, have identified additional rocky exoplanets exhibiting similar tails \citep{haswell2}.
\end{enumerate}

Sections \ref{sec:section2} and \ref{sec:section3} describe the equations and databases used in our general model. Section \ref{sec:section4} describes the $\beta$ ratio results and the velocity boost before ionisation. Section \ref{sec:section5} validates our results by comparing them with previous research, highlighting the improvements made in our model. It addresses the limitations associated with the EUV component of the theoretical model. It also explores three examples in which our model can be effectively applied, providing context for its practical implications. Section \ref{sec:section6} summarises the key findings of our study.


\section{Modelling approach}\label{sec:section2}


\subsection{Calculating $\beta$ ratio}

The radiation pressure coefficient, $\beta$, quantifies the relative influence of radiation pressure and gravity on an individual atom.
Newton's law of gravity defines the gravitational force ($F_{\text{grav}}$) as dependent on the masses of the star and of the atom and their relative distance as
\begin{eqnarray}
    F_{\text{grav}} = \frac{G M m}{r^2}\ ,
    \label{eq:grav}
\end{eqnarray}
where $G$ is the gravitational constant, $M$ is the stellar mass, $m$ the mass of the atom, and $r$ the distance between the star and the atom. To understand the motion of an atom in the vicinity of a star, we must also consider the radiation force ($F_{\text{rad}}$) acting on the individual atom by summing up the impact of the radiation force on the electronic transitions of each atom
\begin{eqnarray}
    F_{\text{rad}} = \frac{1}{8 \pi c^2} \sum_{\text{j<k}} \frac{g_{\text{k}}}{g_{\text{j}}} A_{\text{kj}} \lambda_{\text{jk}}^4  \Phi_{\lambda} \,
    \label{eq:rad}
\end{eqnarray}
where $c$ is the speed of light, $k$ and $j$ are the higher and lower atomic levels, respectively. The terms $g_{j,k}$ are the statistical weights of the atomic levels, $A_{kj}$ is the Einstein coefficient for spontaneous emission from level $k$ to $j$, $\lambda_{jk}$ is the wavelength of the transition and $ \Phi_{\lambda}$ is the emergent stellar flux density at the line centre, scaled to the distance $r$, where the atom is located. Both $F_{\text{grav}}$ and $F_{\text{rad}}$ follow inverse square laws, so their ratio is constant regardless of the distance $r$. Combining equations (\ref{eq:grav}) and (\ref{eq:rad}) gives
\begin{eqnarray}
    \beta = \frac{F_{\text{rad}}}{F_{\text{grav}}} = \frac{1}{8 \pi c^2}  \frac{r^2}{G M m} \sum_{\text{j<k}} \frac{g_{\text{k}}}{g_{\text{j}}} A_{\text{kj}} \lambda_{\text{jk}}^4 \Phi_{\lambda} \, .
    \label{eq:beta}
\end{eqnarray}
For simplicity, we consider regions with low gas temperature and density, where radiative de-excitation dominates over radiative excitation or collisional processes. So most atoms are then in their ground state ($j = 0$), as shown in debris discs by \citet{liseau}, and so we consider only resonance lines to compute $F_{\text{rad}}$. We limit our study to the first three ionisation states of elements from hydrogen to nickel.

\subsubsection{Calculating $\beta$ ratio uncertainties}

The uncertainties in the $\beta$ ratio, presented in \ref{Appendix:tables} Table \ref{table:beta}, were calculated using the following expression:

\begin{eqnarray}
\left( \frac{\delta \beta}{\beta} \right)^2 = 
\left( \frac{\delta \Phi}{\Phi} \right)^2_{\mathrm{cal}} + 
\frac{1}{\beta^2} \sum_i 
\left( \frac{g_{0,i}}{\sum g_0} \right)^2 \beta_i^2 
\left( \frac{\delta A_{i0}}{A_{i0}} \right)^2
    \label{eq:unc}
\end{eqnarray}
where $\delta \beta$ is the uncertainty in the $\beta$ ratio, as listed in Table \ref{table:beta}. Here $\left( \frac{ \delta \Phi}{\Phi} \right)_\mathrm{cal}$ is the fractional uncertainty on the computed stellar flux, $g_0 / \sum g_0$ is the weighting factor accounting for the contribution of multiple ground states. $A_{i0}$ and $\beta_i$ are the Einstein coefficient and $\beta$ ratio corresponding to the $i$-th transition, respectively \citep{fernandez}.

The relative uncertainty values for the stellar flux range from $5$\,\% to $9$\,\% \citep{Bertrone2008}. Stellar fluxes used in our analysis are obtained by scaling the model SEDs to the adopted stellar radius $R_{\odot}$ (Table~\ref{table:param}). The quoted $5$\,\% to $9$\,\% values are the relative uncertainties on the SED normalisation given $R_{\odot}$. Uncertainties for $\beta$ are propagated with Eq.(4) and are conditional on the adopted stellar mass $M_{\odot}$ (Table~\ref{table:param}). To ensure a conservative estimate, the uncertainties of $A_{i0}$ were taken individually for each line from the NIST Atomic Spectra Database. The uncertainties of the Einstein coefficients then spanned from $\leq 0.03$\,\% up to $100$\,\%. The uncertainties on the weighted averages ranged from $0.3$\,\% to $6$\,\%.

\subsection{Velocity boost of neutral atoms}

In the case of an optically thin gas, neutral atoms undergo a "velocity boost" (v$_{\rm ion}$) before ionisation. This boost is determined by the frequency of photon interactions with atoms and the ionisation potential of the atom. The most important factor of this boost is the ionisation rate, defined by the number of ionising photons from the star that strike an atom per unit of time
\begin{eqnarray}
    \Gamma_* = \int_0^{\infty} \frac{\Phi_{\lambda} }{hc/\lambda} \sigma_{\text{ion}} d\lambda\,.
    \label{eq:gamma}
\end{eqnarray}
where $\Gamma_*$ is the ionisation rate per second and $\sigma_{\text{ion}}$ is the ionisation cross section. The velocity boost before ionisation is then given by
\begin{eqnarray}
    v_{\rm ion} \approx \beta \frac{G M}{r^2 \Gamma_*}\,.
    \label{eq:ion}
\end{eqnarray}
The velocity boost is distance independent due to the $r^2$ term in the equation and is relevant only to species that have $\beta > 0.5$. We take the distance as $100$\,AU, because this is the average mid-point in a debris disc. Velocity boost is one way to understand which gas species might form a gas disc and which elements might form a radiation-driven wind \citep{fernandez}.

\section{Input Data}\label{sec:section3}

The stellar fluxes were taken from a collection of {\sc atlas9} flux models\footnote{\url{https://wwwuser.oats.inaf.it/fiorella.castelli/grids/gridp00k2odfnew/fp00k2tab.html}} \citep{atlas,castelli2003_atlas} for stars with effective temperatures $T_{\rm eff} = 4,000 - 19,000$\,K, the surface gravity $\log g$ = 4.5, the metallicity [M/H] as solar and the microturbulence velocity $\nu_{\rm turb}$ = $2.0$\,km\,s$^{-1}$. We extract the stellar mass and radius corresponding to each considered $T_{\rm eff}$ value from the CMD version 3.7 evolutionary tracks\footnote{\url{http://stev.oapd.inaf.it/cgi-bin/cmd}} \citep{cmd1,cmd2, cmd3, cmd4, cmd5, cmd6}, assuming solar composition. A summary of the adopted stellar parameters is provided in \ref{Appendix:tables}, Table~\ref{table:param}. We focus on early-type stars at an age of $20$\,Myr, as our study focuses, for comparison reasons, more on gas-rich, young debris discs. As we see in the Section \ref{sec:section5}, the age does not matter much.

The input line parameters for computing the radiation pressure were obtained from the NIST Atomic Spectra Database version 5.11\footnote{\url{https://www.nist.gov/pml/atomic-spectra-database}}, where H to Ni atoms are considered with their first three ionisation states \citep{nist}. Our analysis considers only observed wavelengths, except for \ion{He}{II}, \ion{Ni}{II}, where Ritz wavelengths were used due to the lack of observed wavelengths. We focus only on allowed transitions, labelled in the database as "E1". We did not include \ion{Ni}{III} which has only forbidden transitions from its ground state. The examined spectral region spans the $100<\lambda<5,000$\,nm wavelength range since resonance lines below $100$\,nm do not significantly affect the $\beta$ ratio. Because the model under consideration, {\sc atlas9}, is not good at replicating the ultraviolet region of the spectral energy density (SED)). For further analysis, we also incorporate corrected EUV flux and for that, we also take into account the spectral region below $100$\,nm into our calculations, see Section \ref{sec:betasun}.

Input data for the ionisation cross sections were obtained from the NIST XCOM photon cross sections database\footnote{\url{https://www.nist.gov/pml/xcom-photon-cross-sections-database}} \citep{xcom}. This comprehensive repository contains experimentally derived cross sections for photon interactions with atomic and molecular species. 
We used the total attenuation coefficients from this database, which include all interactions such as photo-ionisation and coherent scattering. These values were then multiplied by the atomic mass of each element to obtain a reasonable upper limit for the ionisation cross-sections.


\begin{figure*}
\centerline{\includegraphics[width=0.96\textwidth]{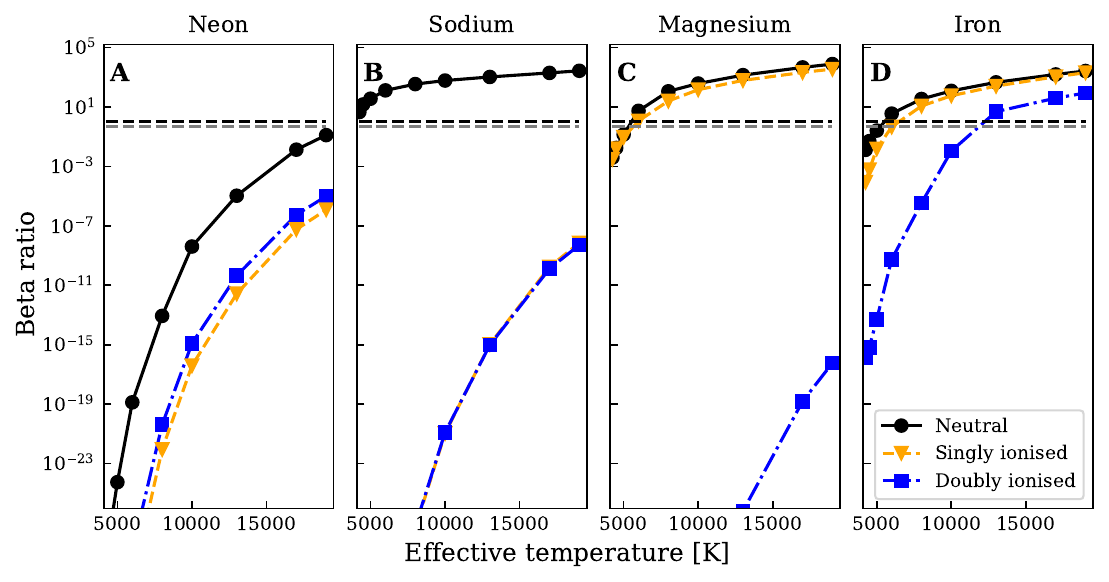}}
    \caption{$\beta$ ratios of species representative of selected groups as a function of stellar effective temperature. Each panel shows three ionisation states for each species. The black dashed line ($\beta = 1$) indicates the equilibrium condition, where the radiative and gravitational forces acting on the atoms are balanced and the grey dashed line ($\beta = 0.5$) indicates the $\beta$ value above which unhindered gas, when released from circular orbit, starts to migrate outwards.}
    \label{fig:neut}
\end{figure*}

\begin{figure*}
\centerline{\includegraphics[width=0.96\textwidth]{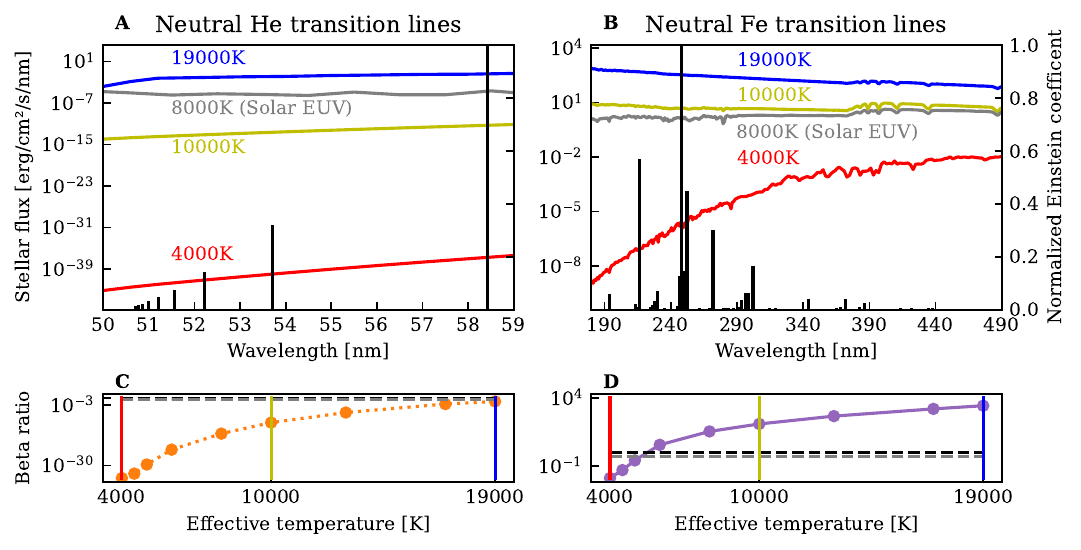}}
    \caption{Panels (A) and (B) show the strengths of neutral helium and iron transitions compared to the fluxes emitted by stars with various effective temperatures. The complementary panels (C) and (D) show $\beta$ ratios of neutral helium and iron, respectively. The black dashed line marks the equilibrium between gravitational and radiation forces. Panels highlight the limitations of the {\sc atlas9} models in the ultraviolet regime compared to the $T_{\rm eff} = 8,000$\,K model, which is corrected with Solar EUV flux, as explained in Section \ref{sec:betapic}. Effective temperatures of $4,000$\,K, $10,000$\,K and $19,000$\,K are indicated above the flux curves in the upper panels and from left to right in the lower panels and are distinguished by different colours.
}
    \label{fig:helium}
\end{figure*}


\section{Results}\label{sec:section4}

Using the above methods and data, we get the following results for the $\beta$ ratio and the velocity boost of different atomic ionisation states. The code used in this work is publicly available on GitHub\footnote{\url{https://github.com/Drannu/Betaratio}}. The results are available as machine-readable tables at the CDS via anonymous ftp to \texttt{cdsarc.cds.unistra.fr} (130.79.128.5)\footnote{\url{https://cdsarc.cds.unistra.fr/viz-bin/cat/J/AN}}.


\subsection{$\beta$ ratio}

Figure~\ref{fig:neut} shows the $\beta$ ratios obtained for selected neutral (black circles), singly ionised (orange triangles) and doubly ionised (blue squares) species as a function of the stellar effective temperature. For all species, $\beta$ increases with increasing stellar effective temperature. At the lowest $T_{\rm eff}$ values, $\beta$ rises steeply at about $2\,$dex per $1,000\,$K,  and then flattens typically at around $10,000$--$12,000$\,K to about $1\,$dex per $4,000\,$K. 

We grouped similarly behaving species together to better clarify their behaviour across different ionisation states. These groups are: noble gases (i.e. He, Ne, Ar), IA group elements (i.e. Li, Na, K), IIA group elements (i.e. Be, Mg, Ca) and Fe-peak elements (i.e. Cr, Mn, Fe, Co, Ni). The chosen representative species for each group are: Neon (representative of noble gases), Sodium (representative of IA group), Magnesium (representative of IIA group) and Iron (representative of Fe-peak group). The results for three ionisation states of the representative elements are summarised in \ref{appendix:figures} Figures \ref{fig:allneut}, \ref{fig:allsinglyionised} and \ref{fig:alldoublyionised}. In the doubly ionised case, \ion{Li}{III} and \ion{Be}{III} are omitted due to a lack of available atomic data.

The $\beta$ ratio exhibits a flatter trend at higher effective temperatures, mainly due to the flux modelled by {\sc atlas9} and the specific locations of the strongest transition lines for various chemical species. This effect is demonstrated in Figure~\ref{fig:helium}. The upper panels (A) and (B) illustrate the flux variation across different effective temperatures, also showing the positions of transitions for neutral helium and iron. Notably, the transition with the largest Einstein coefficient plays the most significant role in shaping the $\beta$ ratio. The lower panels (C) and (D) present the corresponding $\beta$ ratio for these neutral species. A comparison of the flux at the strongest transition line across featured effective temperatures reveals that as the flux magnitudes converge at higher temperatures, the $\beta$ ratio becomes increasingly flat.

\subsubsection{Neutral atoms}

Figure \ref{fig:neut} shows the $\beta$ ratios as a function of effective temperature for representative neutral atoms. Due to the small number of resonance lines, noble gases could have $\beta$ values much higher than one only for stars of high temperature. 

In particular, \ion{He}{I} and \ion{Ne}{I} will not become unbound from the star due to radiation pressure in the star in the entire effective temperature range explored in this study ($4,500$--$19,000$\,K). Remarkably, helium has one of the lowest beta values as a neutral atom and as singly and doubly ionised states compared to other chemical elements. 

The low $\beta$ values obtained for He, particularly in comparison to Fe-peak elements, result from a combination of the small number of resonance lines, the weakness of these lines, and the low stellar flux at the wavelengths of the lines, as illustrated in Figure\,\ref{fig:helium}. In general, for the Fe-peak elements, we find that $\beta$ values are above one if the stellar $T_{\rm eff}$ is over $7,000\,$K. For IA group elements, the $\beta$ ratios are always above one in the explored temperature range.

\subsubsection{Singly ionised atoms}

Figure \ref{fig:neut} illustrates the $\beta$ ratios for the representative singly ionised ions. For these ions, we find that $\beta$ values increase with stellar effective temperature, similar to neutral species, though generally lower than neutral species. 
Remarkably, we find that the $\beta$ values for IA group elements, which in the neutral form have one electron in the outer shell, decrease dramatically when they are singly ionised and remain below one throughout the entire considered $T_{\rm eff}$ range. This phenomenon suggests that these elements, once ionised, would be less subjected to the stellar radiation pressure and thus will be eventually accreted. However, singly ionised species are also affected by the magnetic field of the host star, but we do not consider this effect in this work.


\subsubsection{Doubly ionised atoms}

Figure \ref{fig:neut} presents the $\beta$ ratios for representative doubly ionised species. As before, we find $\beta$ values that generally increase with increasing $T_{\rm eff}$, followed by a plateau, but with even smaller values compared to the case of the singly ionised species. Interestingly, we find no species with a $\beta$ value more significant than one for stars cooler than $8,000$\,K, implying that for solar-like stars, doubly ionised species are pulled towards the star. However, one would then have to consider the impact of stellar magnetic field and wind, which goes beyond the scope of this work.

\begin{figure}[b]
\centerline{\includegraphics[width=0.50\textwidth]{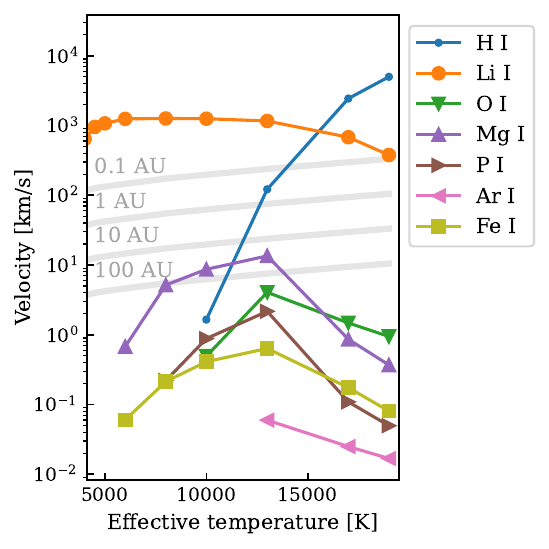}}
    \caption{Velocity boosts of selected neutral atoms with \(\beta > 0.5\) as a function of stellar effective temperature. The atoms are arranged across the panels in descending order of their \(\beta\) values at \(T_{\mathrm{eff}} = 19,000~\mathrm{K}\). Shown in grey are the escape velocities at distances of \(0.1\), \(1\), \(10\), and \(100~\mathrm{AU}\).}
        \label{fig:vel}
    \label{fig:vel}
\end{figure}

\subsection{Velocities}

Figure \ref{fig:vel} presents the velocity boost obtained for $7$ selected neutral atoms when $\beta > 0.5$. The computed velocities range from about $0.01$ to $100$\,km/s for most neutral atoms, with the actual values being mainly controlled by the ionisation cross section. All of the velocity boosts of chosen species are given in \ref{appendix:figures} Figure \ref{fig:allvel}. 

The theoretical velocity boost will only result in the loss of the particle if $v_{\rm ion}$ exceeds the local escape velocity from the star, $v_{\rm esc}$. Otherwise, a collisionless particle with $v_{\rm ion}<v_{\rm esc}$ will be pushed to an eccentric orbit prior to being ionised. We have therefore included the escape velocities at $0.1$, $1$, $10$ and $100$\,AU in Figures\,\ref{fig:vel} and \ref{fig:scenvel} using the following equation:
\begin{eqnarray}
    v_{\rm esc} = \sqrt{\frac{2 G M}{r_g}}\,,
    \label{eq:escapevelocity}
\end{eqnarray}
where $G$ is the gravitational constant, $M$ is the star's mass, and $r_g$ is the distance between the star's center and the circumstellar gas's location.

The velocity boost trend shows that circumstellar gas moves away from the star at higher effective temperatures at faster velocities. Most heavy elements remain bound to the system at distances of $10$ AU or less. However, species like \ion{Be}{I}, \ion{O}{I}, \ion{Al}{I}, and \ion{V}{I} need the object to be within $1$ AU to stay in orbit, while \ion{H}{I} and \ion{B}{I} require an even closer distance, less than $0.01$ AU. Neutral hydrogen is most likely to escape, especially at high effective temperatures.

Interestingly, IA group elements behave differently. They are more likely to escape at lower temperatures, making these elements easier to lose from the system when the star is cooler.


\section{Discussion}\label{sec:section5}

Our model can be used in many different environments. Our main aim was to describe the evaporation from exoplanets and the gas in debris discs to determine the composition of material moving towards and away from the star. We can also use this model on circumstellar, accretion, and decretion discs. Situations where the corpuscular stellar wind is dominant, which may be the case for decretion discs, are not accurately covered by our model since we focus on calculating the radiation pressure \citep{kral}. We can also look at a star passing through interstellar material, creating bow shock in gas and dust, where both radiation and stellar wind play an important role \citep{ochsendorf}. Our model can also describe the behaviour of the evaporated material from disrupted planets and possibly the material of a circumplanetary disc. We can also use this model for smaller objects, such as the evaporation of exocomet gas \citep{exocomet1, exocomet2}. 


\subsection{Comparison with previous works}\label{sec:betapic}

The $\beta$\,Pictoris system is centered around an A5V star that hosts a gas-rich debris disc. The gas is dominated by C, O and other non-hydrogen elements and is thought to be of secondary origin, i.e., released from icy planetesimals. The primary braking agents of the gas in the disc are believed to be carbon and oxygen. This opinion is based on the $\beta$ values and velocity boosts of neutral atoms calculated by \citet{fernandez}. These values were computed for elements from hydrogen to nickel, using a {\sc phoenix} model spectrum to reproduce the stellar flux. They found that heavily ionised species in the debris disc are predominantly influenced by radiation, with velocity boosts of short-lived neutral atoms aligning closely with observational constraints. From the $\beta$ ratio values, they concluded that neutral carbon and oxygen do not experience significant radiation pressure from the star and can thus accumulate over time \citep{fernandez}. 

One of the stellar temperatures considered in this work has been chosen to roughly match that of $\beta$\,Pictoris and thus allow a comparison with the results of \citet{fernandez}. The only difference is that, in our case, the stellar flux was computed assuming a stellar mass of $1.75$\,M$_\odot$, effective temperature as $8,073$\,K and stellar radius of $1.53$\,R$_\odot$ \citep{DiFolco2004}, whereas \citet{fernandez} used the observed flux directly. The main difference from our previous calculations lies in the assumed stellar radius: previously, a value of $1.03$\,R$_\odot$ was adopted from a generic isochrone model, which yields a smaller stellar radius than that used in the present calculations (see Table~\ref{table:param} for the adopted stellar parameters).

\begin{figure}
\centerline{\includegraphics[width=0.5\textwidth]{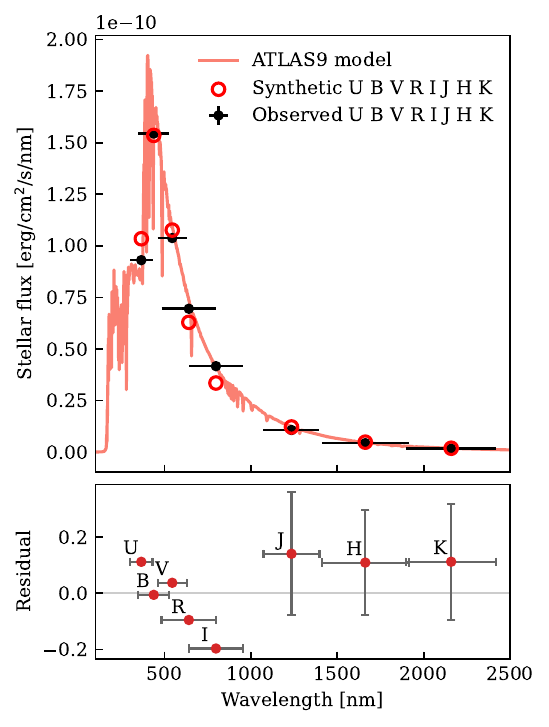}}
    \caption{Top: Comparison between the observed (black dots) and synthetic (red circles) photometric fluxes of $\beta$~Pictoris in the Johnson ($U$, $B$, $V$, $R$, $I$) and 2MASS ($J$, $H$, and $K$) bands. The red solid line shows the \textsc{atlas9} model. Bottom: Fractional residuals $(F_{\mathrm{syn}} - F_{\mathrm{obs}})/F_{\mathrm{obs}}$ between observed and synthetic fluxes.}
    \label{fig:comp_proof}
\end{figure}

Therefore, to enable an adequate comparison, we derived the stellar radius to be used for computing the $\beta$ ratios by matching the \textsc{atlas9} synthetic spectral energy distribution with broadband Johnson $U$, $B$, $V$, $R$, $I$ \citep{ducati2002} and 2MASS $J$, $H$, $K$ \citep{Squicciarini2025} photometry (see Fig.~\ref{fig:comp_proof}). The synthetic magnitudes were computed by integrating the product of the \textsc{atlas9} spectrum and the corresponding filter transmission curves over wavelength.

The best-fitting \textsc{ATLAS9} model reproduces the observed spectral energy distribution (SED) of $\beta$~Pictoris for a stellar radius of $R_\star = 1.53 $\,R$_\odot$. The overall agreement between the observed and synthetic fluxes across the optical and near-infrared bands indicates that the flux calibration and adopted stellar parameters are generally consistent.

A systematic trend is visible in the residuals of the Johnson $U$, $B$, $V$, $R$, and $I$ bands, while the near-infrared $J$, $H$, $K$ points show $+1\sigma$ overestimations. If the discrepancy were confined to the optical region, it could be attributed to a mild interstellar reddening effect. However, the $J$, $H$ and $K$ fluxes are also slightly overestimated, suggesting that the adopted effective temperature may be somewhat inaccurate. In particular, the model appears to be marginally too hot, leading to an excess flux in the blue and insufficient flux in the red spectral region of Johnson bands. A reduction of approximately $150$\,K in $T_{\mathrm{eff}}$ would likely improve the overall fit.

\begin{figure}
    \centerline{\includegraphics[width=0.8\linewidth]{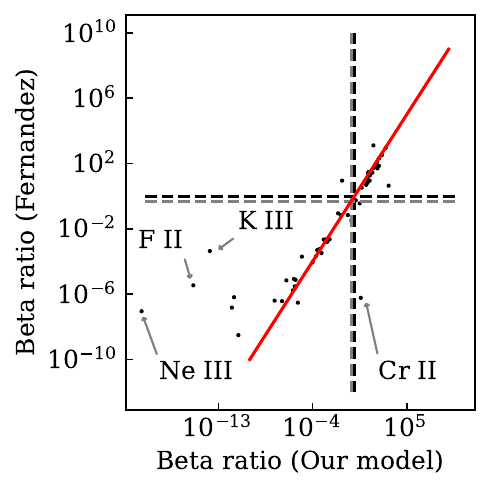}}
    \caption{Comparison between the $\beta$ ratios obtained for the star $\beta$ Pictoris in this work and the results from \citet{fernandez}. The red line represents the equality between the two sets of values. The grey line represents the $\beta$ value above which unhindered gas, when released from circular orbit, starts to migrate outwards.}
    \label{fig:compbeta}
\end{figure}

\begin{figure}
    \centerline{\includegraphics[width=0.8\linewidth]{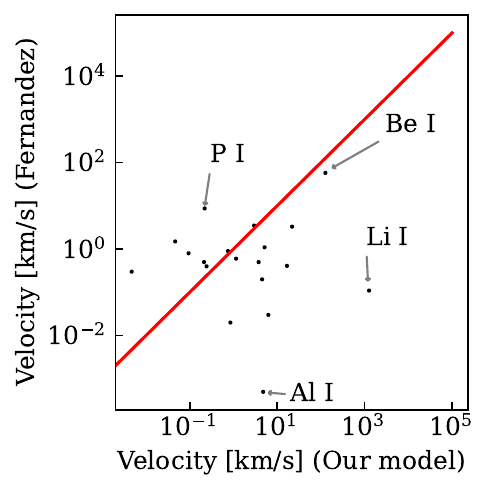}}
    \caption{Comparison between the velocity boosts obtained in this work and the results from \citet{fernandez}. The red line represents the equality between the two sets of values.}
    \label{fig:compbeta2}
\end{figure}

\begin{figure*}[t]
    \centerline{\includegraphics[width=0.98\linewidth]{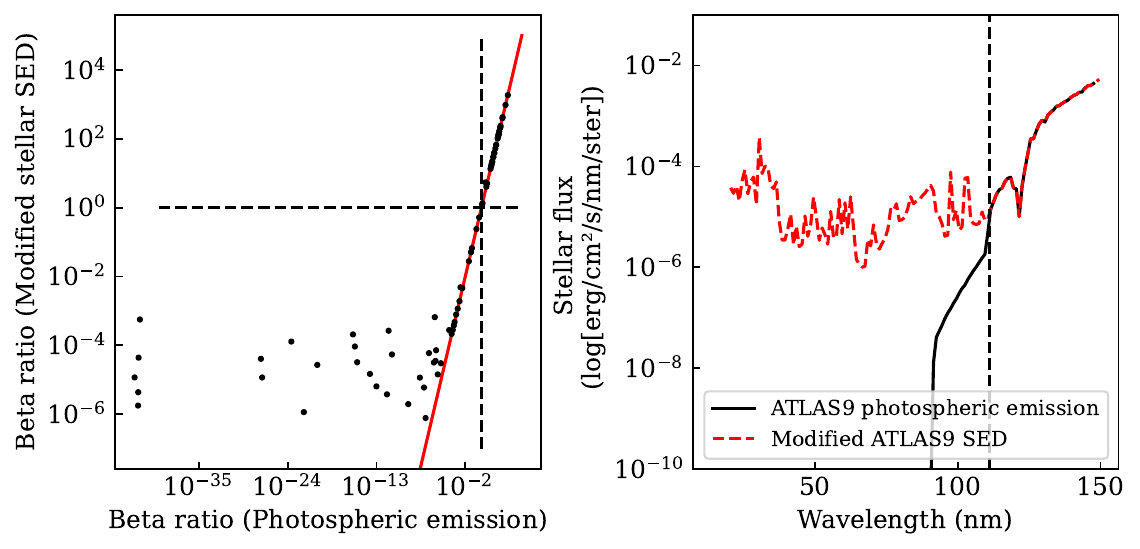}}
    \caption{Panel (A) shows the comparison between $\beta$ ratios obtained in the case of $\beta$\,Pic including (y-axis) or excluding (x-axis) the stellar chromospheric/coronal emission. The red line represents the equality between the two sets of values. The grey line represents the $\beta$ value above which unhindered gas, when released from circular orbit, starts to migrate outwards. Panel (B) shows the spectral energy distribution considering the photospheric emission (black) and Solar modified {\sc atlas9} model (red). To include chromospheric/coronal emission in the stellar SED, we substitute the photospheric emission (black) in the EUV and X-ray bands with the scaled solar SED (red). Black vertical line shows the wavelength below which the substitution occurred.}
    \label{fig:compflux}
\end{figure*}

We find that the $\beta$ ratio values reported by \citet{fernandez} are, on average, slightly lower than ours, which can be attributed to the different stellar flux models used. Comparisons of the $\beta$ values and velocity boosts are given in Figures \ref{fig:compbeta} and \ref{fig:compbeta2}, respectively. Numerical comparison of $\beta$ ratios is given in the \ref{Appendix:tables} Table \ref{tab:comp fernandez} and for velocity boosts in the \ref{Appendix:tables} Table \ref{tab:comp Fernandez velocity}. Numerous works used the results of \citet{fernandez}. For example, \citet{czechowski} calculated the column density for \ion{Ca}{II} using the $\beta$ value from their results, finding it is smaller than the observed values. Had they employed our $\beta$ value, they would have obtained a column density for \ion{Ca}{II} more consistent with observations. 

When examining the braking agents of the gas in the $\beta$ Pictoris disc, we find that the $\beta$ value for \ion{C}{II} and \ion{C}{III} is of the same order of magnitude, while \ion{C}{I} is one order of magnitude higher than the values obtained by \citet{fernandez}. For \ion{O}{I}, we find a $\beta$ value is in the same order of magnitude as \citet{fernandez}. For \ion{O}{II}, we obtain a smaller value by about 3 orders of magnitude, while for \ion{O}{III} we get about 6 orders of magnitude lower value. From these results, we can say that carbon is a better braking agent than was found earlier. However, iron-peak elements are more strongly affected by radiation pressure.

\begin{table}
\centering
\caption{Comparison of our calculated $\beta$ ratios to the theoretical model spectrum with the observational constraints from the HST spectra as reported by \citet{Lagrange1998, Lagrange1996}, references as L96 and L98 in the table.}
\label{tab:HSTvsUs}
\renewcommand{\arraystretch}{1.1}
\begin{tabularx}{\columnwidth}{l|YYY}
\hline
Ion             & Our model & HST data L96 & HST data L98           \\ \hline
$\ion{C}{I}$    & $0.52$    &              & $0.011 - 0.095$        \\
$\ion{Mg}{I}$   & $130$     &              & $47 - 82$              \\
$\ion{Mg}{II}$  & $29$      & $3.47$       & $0 - 4.96$             \\
$\ion{Al}{I}$   & $66$      &              & $\geq 7.3 - \geq 10.3$ \\
$\ion{Al}{II}$  & $3.2$     & $0.394$      & $0 - 0.56$             \\
$\ion{Al}{III}$ & $16.7$    & $10.9$       & $9.8 - 15.6$           \\
$\ion{Si}{I}$   & $15.2$    &              & $\geq 8.8 - 10.1$      \\
$\ion{Si}{II}$  & $0.07$    &              & $0.12 - 0.15$          \\
$\ion{S}{I}$    & $1.4$     &              & $1.46 - 1.52$          \\
$\ion{Ca}{II}$  & $160$     &              & $4.4 - 35$             \\
$\ion{Cr}{II}$  & $4.0$     &              & $2.5 - 3.36$           \\
$\ion{Mn}{II}$  & $18.4$    &              & $\geq 2.9 - 11.0$      \\
$\ion{Fe}{I}$   & $39.6$    &              & $14.7 - 19.1$          \\
$\ion{Fe}{II}$  & $13.4$    &              & $0.44 - 4.87$          \\
$\ion{Ni}{I}$   & $20.8$    &              & $\geq 0.15 - \geq 0.6$ \\
$\ion{Ni}{II}$  & $0.2$     &              & $1.38 - 1.39$          \\ \hline
\end{tabularx}
\end{table}

In Table~\ref{tab:HSTvsUs}, we compare the calculated $\beta$ ratios from our model with the Hubble Space Telescope (HST) constrained values. Overall, our $\beta$ estimates are systematically higher than the constraints of \citet{Lagrange1996, Lagrange1998} for ions with strong resonance transitions. The largest discrepancies are for $\ion{Mg}{I}$, $\ion{Mg}{II}$, $\ion{Ca}{II}$, $\ion{Fe}{I}$, and $\ion{Fe}{II}$, where our values exceed the HST inferred ranges by factors of a few up to an order of magnitude. By contrast, $\ion{S}{I}$ and $\ion{Al}{III}$ agree within $\sim10-40$\%, and $\ion{Cr}{II}$ is close to the observed range. Our prediction for $\ion{Si}{II}$ lies slightly below the observed interval, and $\ion{Ni}{II}$ is lower by a factor of $\sim$ 1. For species with only lower limits reported, our $\beta$ values comfortably exceed those limits. This overall bias towards higher $\beta$ may indicate that the adopted stellar ultraviolet fluxes and/or line strengths are overestimated, or that radiative transfer effects are not yet fully captured.

Similarly to \citet{fernandez}, \citet{youngblood} computed the $\beta$ ratios for various ions, including \ion{Fe}{II}, \ion{Si}{II}, \ion{Mg}{II}, \ion{S}{I}, \ion{S}{II}, \ion{Al}{II}, \ion{C}{I}, \ion{O}{I}, and \ion{C}{II}, for stellar effective temperatures ranging from $8,000$ to $13,000$\,K. They found that their $\beta$ ratio values are $1.3$ to $190$ times higher than found by \citet{fernandez}. These differences were attributed to updated synthetic spectra, improved atomic data, and variations in stellar flux densities. Compared to our $\beta$ ratios, the values reported by \citet{youngblood} are within the uncertainty range. 

Our model does not include rotational broadening effects on the flux distribution in the line profile, a factor considered in the work of \citet{fernandez}. The {\sc phoenix} model, that \citet{fernandez} used, produces high-resolution spectra, where rotational broadening is important. In contrast, the {\sc atlas9} model, that we use in this article, gives low-resolution spectra, so rotational broadening has little effect. What we do expect to make a difference, however, is that the {\sc atlas9} model assumes local thermodynamic equilibrium (LTE), whereas {\sc phoenix} considers non-LTE effects. The outcome is that the EUV photospheric emission of the spectrum computed assuming LTE (i.e. {\sc atlas9}) is 3-6 magnitude lower than that obtained considering non-LTE effects (i.e. {\sc phoenix}) \citep{Schaerer1997A}. 


\begin{figure*}
    \centerline{\includegraphics[width=0.98\textwidth]{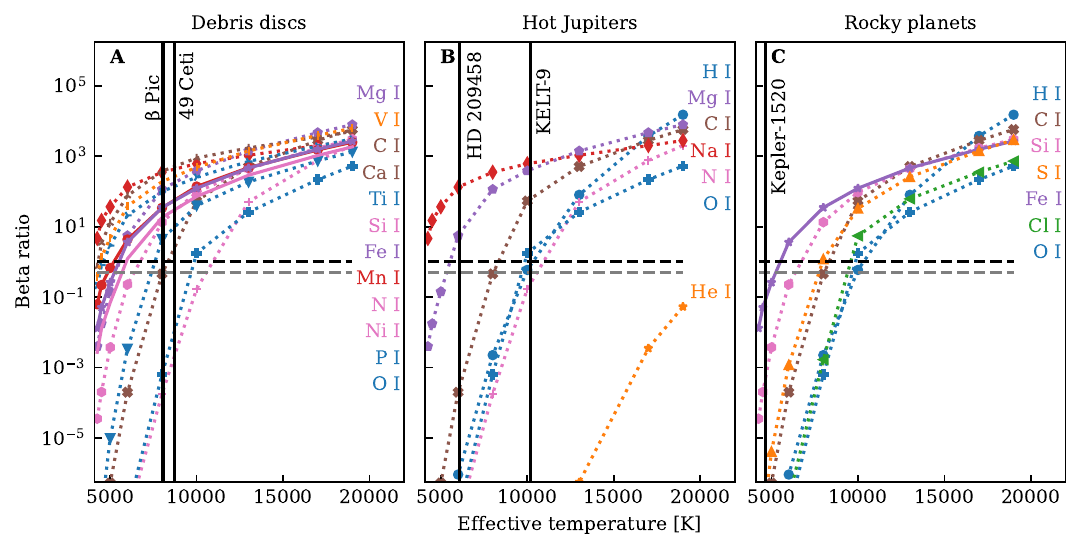}}
    \caption{Three distinct evaporation scenarios are depicted in terms of the $\beta$ ratio as a function of the star's effective temperature: (A) Secondary gas debris discs, (B) hot Jupiters, and (C) rocky planets. The atoms are arranged across the panels in descending order of their \(\beta\) values at T$_{\rm eff}$ = $19,000$ \,K. The black dashed line represents the equilibrium point of gravitational and radiation forces. The grey dashed line represents $\beta$ value above which the unhindered gas, when released from circular orbit, starts to migrate outwards.}
    \label{fig:scenbeta}
\end{figure*}

\begin{figure*}
  \centerline{\includegraphics[width=0.98\textwidth]{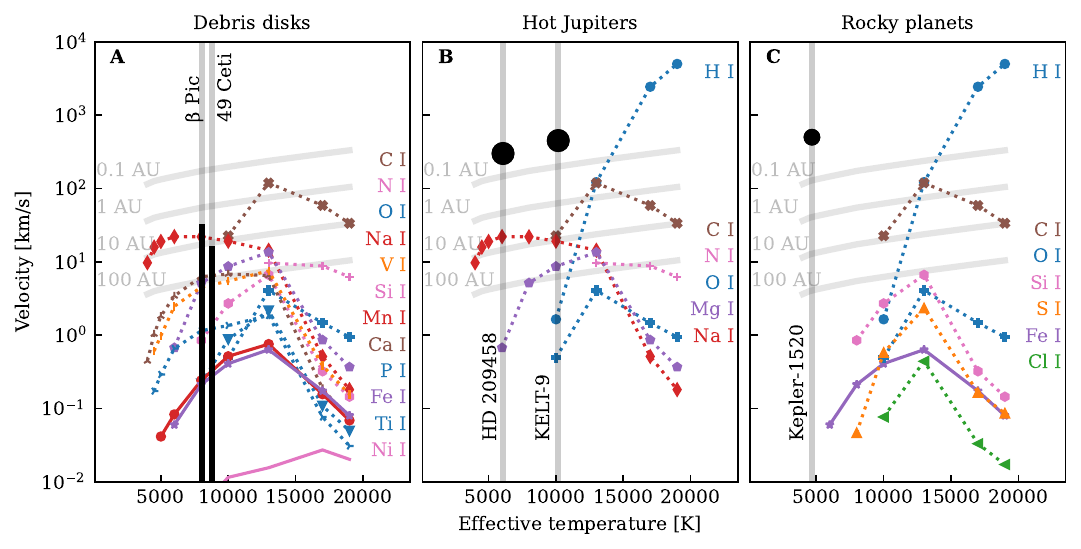}}
  \caption{Velocity boosts prior to ionisation ($v_{\rm ion}$) as a function of the star's effective temperature for atomic species that are relevant for three scenarios of atomic circumstellar gas: (A) Secondary gas-rich debris discs, (B) hot Jupiters, and (C) rocky planets. The atoms are arranged across the panels in descending order of their $\beta$ values at T$_{\rm eff}$ = $19,000$\,K. Shown in horizontal grey lines are the escape velocities at distances of $0.1$, $1$, $10$, and $100\,\mathrm{AU}$. Vertical grey lines represent the effective temperatures of a selection of prominent systems within each category, with the overlaid black line or dot showing the escape velocity of the chosen systems.}
  \label{fig:scenvel}
\end{figure*}


\subsection{The impact of a more realistic EUV stellar emission}\label{sec:betasun}

Several atoms exhibit strong resonance lines in the EUV band, where stellar emission can be significant due to the presence of a chromosphere. While these lines are well characterised for solar-like stars, the presence and strength of EUV emission in intermediate-mass stars (spectral types A and F) remain poorly understood. Observations indicate that stars with effective temperatures up to approximately $8400$\,K do show emissions typically associated with chromospheric and/or coronal activity, suggesting that they may also possess substantial EUV emission, potentially several times stronger than solar levels \citep[see, e.g.][]{fossati}.

The synthetic stellar fluxes provided by {\sc atlas9} are purely photospheric and thus do not include any contribution from the chromosphere and/or corona, which would significantly increase the emission, particularly in the EUV band. Therefore, for stars cooler than about $8,400$\,K, the $\beta$ values reported in Section \ref{sec:section4} might be underestimated. To assess the impact of more realistic stellar EUV emission, we focus on the case of $\beta$\,Pictoris, which has an effective temperature of approximately $8,000$\,K. \citet{Wu2025} concluded that the observed strength of the \ion{Ar}{II} line in $\beta$\,Pictoris can only be explained if the chromospheric and/or coronal EUV emission is comparable to that of the modern Sun. However, no observations of $\beta$\,Pictoris exist to constrain the EUV emission, introducing significant uncertainty. To account for this, we replaced the stellar EUV emission predicted by {\sc atlas9} with that of the quiet Sun \citep{claire2012_sun}, scaled to match the stellar radius of $\beta$\,Pictoris ($1.53$ R$_\odot$) and further multiplied by a factor of three, following the suggestion by \citet{fossati} to reconcile the solar EUV flux with that observed for Procyon by the EUVE mission \citep{craig}.

Including a chromospheric and/or coronal EUV emission has significantly impacted the $\beta$ ratio values, particularly for the species that showed the lowest values without EUV emission (see Figure \ref{fig:compflux}). Although for some species, the difference amounts to several orders of magnitude (up to a factor $10^{35}$), for none of them, the $\beta$ values increase from below to above unity. We verified the same behaviour for 49\,Ceti, HD\,209458, KELT-9, and Kepler-1520: incorporating the scaled quiet-Sun EUV slightly increases $\beta$ but does not increase any species above unity. So, accounting for the presence of a stellar chromospheric and/or coronal emission should not drastically alter the fate of the particles, but rather alter their velocity.

Future research should address these limitations by examining ion interactions in various environments. Including a stellar wind in a model that accounts for the $\beta$ ratio is important because a stellar wind exerts an additional outward force on particles near the star. This wind affects the flow and distribution of material, potentially altering the dynamics that would otherwise be influenced only by gravity and radiation. Investigating the role of magnetic fields in influencing atomic dynamics could also provide valuable insights.

\subsection{Gas-rich debris disc}\label{disc}

The scenario of a gas-rich debris disc around a range of stars with different effective temperatures is shown in Figure \ref{fig:scenbeta} panel (A). The neutral species were chosen from \citet{sgdd1, sgdd2, brandeker}, representing typical neutral atoms found in debris discs. Also included in the scenario are two real stars with debris discs -- $\beta$\,Pictoris and $49$\,Ceti. These stars have very similar effective temperatures, so they mostly exhibit the same species drifting toward and being pushed away from the star. What we mean by drifting is that a single atom with low radiation pressure force will stay on a stable Keplerian orbit, but depending on the density of the disc viscous spreading may allow the atoms to drift towards the star. We assume in this paper that the disc is thin. The distances from the star to the inner and outer edges of the debris discs around $\beta$\,Pictoris and $49$\,Ceti were taken from \citet{millar} and \citet{Wahhaj}, respectively.

Also, the velocity boost is shown for this scenario before ionising the selected species in panel (A) of Figure \ref{fig:scenvel}. In this specific case, the velocity boost of most atoms is averaging around $0.5$\,km/s.

In the case of $\beta$\,Pictoris debris disc, the gas in the disc is observed to contain H, C, N, O, Na, Mg, Al, Si, P, S, Ca, Cr, Mn, Ar, Fe, and Ni \citep{roberge, Wu2025}. \citet{roberge} report the presence of a gas braking mechanism in the disc, identifying \ion{C}{II} and \ion{C}{III} as the primary braking agents due to their detected absorption lines and their overabundance relative to solar values \citep[see also][]{fernandez} Similarly, in the case of $49$\,Ceti, carbon has also been identified as the key braking agent \citep{roberge2014}. According to our model results, carbon, with its ionisation states, is not significantly affected by stellar radiation pressure and is therefore expected to be stable or inward-drifting under the influence of gravity. Given its overabundance in the disc \citep{fernandez}, carbon can act as a drag force, slowing down other species or preventing them from being expelled from the system.



\subsection{Evaporating hot Jupiter}\label{jupiter}

The following scenario describes what happens to the selected chemical species \citep{vidal-madjar, ehj2, ehj3, ehj4} after leaving the atmosphere of a hot Jupiter through thermal evaporation. This scenario is shown in Figure\,\ref{fig:scenbeta} panel (B) with two well-studied stars hosting hot Jupiters -- KELT-9 and HD\,209458. 

In the case of HD\,209458, our results show that the only species experiencing sufficient radiation pressure to be pushed away from the star are metals. Specifically, elements from groups IA (except hydrogen), IIA, IIB, IVB, VB, VIB, VIIB, VIIIB, as well as aluminium. In contrast, non-metals such as hydrogen, boron, and elements from groups IVA, VA, VIA, and VIIA exhibit $\beta$ values below one. Therefore, their orbits are expected to be stable or inward-drifting around the star. However, we remind that HD\,209458 is likely to have a wind that significantly affects the motion of atoms in the vicinity of the planet \citet{Shaikhislamov}. For the hotter star KELT-9, the only species found to be gravitationally bound (i.e., with $\beta < 0.5$ for material released from a circular robit) are phosphorus, nitrogen, and helium.

The velocity boost of neutral atoms before ionisation for the general scenario of hot Jupiters, specifically for these two well-known systems, is given in Figure\,\ref{fig:scenvel}, panel (B). In the case of HD\,209458, among the considered species, only \ion{Na}{i} gets pushed away from the star before ionisation. The attained velocity boost, $\sim 0.1\,$km\,s$^{-1}$, is well below the escape velocity, and since \ion{Na}{ii} has $\beta<0.5$, we can conclude sodium will remain stably orbiting the star. For KELT-9, the velocity boost is relevant for \ion{O}{i}, \ion{Mg}{i}, and \ion{Na}{i}. All of them get pushed to low velocities, $v_{\rm ion}\lesssim 0.4\,$km\,s$^{-1}$. Similarly to Na around HD\,209458, we can conclude that all these atomic species have orbits that are stable to radiation pressure after being evaporated from hot Jupiter.


\subsection{Evaporating rocky planet}\label{rocky}

In this scenario, we look at specific species found in Earth-like rocky planet atmospheres and their behaviour when they evaporate from the atmosphere. 

The scenario with an example of star Kepler-1520 with a rocky planet is shown in Figure \ref{fig:scenbeta} panel (C). Chemical species were chosen from \citet{erp}. The rocky planet around Kepler-1520, with an Earth-like atmosphere, has been observed to contain non-metals and iron. The $\beta$ ratio is below one for all chemical species, even for iron.

Figure \ref{fig:scenvel} panel (C) shows the velocity boost of the neutral atoms chosen before ionisation. In the Kepler-1520 case, all atoms are moving towards the star. So, from this model, we can predict the presence of a gas torus around the star \citep{campos}.


\section{Conclusions}\label{sec:section6}

This study presents a detailed model for analysing the behaviour of atomic particles under the influence of stellar gravity and radiation pressure. We consider the movement of neutral, ionised, and doubly ionised atoms, including various factors such as stellar types, gravity, and radiation pressure. By examining these interactions, the study offers insights into the dynamics of exoplanetary systems, debris discs and, more generally, any situation where tenuous circumstellar gas is present. Here is a list of our key findings:
\begin{enumerate}[i]
        \item Different atoms and ions exhibit varying motions relative to the star at specific effective temperatures, particularly for A- and F-type stars. For instance, neutral magnesium can orbit stably around the star if the star's effective temperature is below $8,000$\,K, but if the temperature exceeds this threshold, it will be pushed away from the star. The velocity boost experienced by neutral magnesium due to radiation pressure, relative to the escape velocity, suggests that it must be located at a distance greater than $5\,\mathrm{AU}$ in order to escape the system within the given range of effective temperatures.
        
        \item By examining the motion of atoms before and after ionisation, we observe a clear trend: more highly ionised atoms are less affected by radiation pressure as shown in Figure \ref{fig:neut}. This likely reflects the shift of their strongest resonance transitions into the FUV/EUV, where stars with $T_{\mathrm{eff}}\lesssim 10{,}000$\,K, have the photon flux intrinsically low. Consistent with this, applying our EUV flux correction has a negligible impact on the inferred dynamics: the dynamics of the ions remain essentially unchanged for such stars.
        
        \item The effect radiation pressure has on atomic species depends on the configuration of their outermost electrons. Like noble gases, atoms with closed electron shells are more stable and interact weakly with stellar radiation. In contrast, atoms with unclosed shells, such as metals with 1–3 outer electrons, interact more strongly. For instance, neutral sodium, with one outer electron, strongly interacts with stellar radiation. However, once ionised and its outer shell is closed, its interaction weakens (Figure \ref{fig:neut}).
        
        \item We can see from our results, that chromosphere and/or corona influence on radiation coefficients, for stars that have effective temperature below $8,000$\,K, is not very significant.The species that have the strongest line in the EUV region have $\beta$ ratio rise by many magnitudes, but not enough to alter the physics of the behaviour. As a counterargument, \citet{youngblood} has theorised that EUV flux should be significant. That may be true for more hotter stars.
        
        \item The most probable chemical species that can contaminate the stellar photospheres of early-type stars, $6,000 < T_{\rm eff}< 10,000\,$K, are noble gases, singly and doubly ionised non-metals and most of the doubly ionised metals.
        
        \item Previous research by \citet{jermyn} indicates that the detection of stellar photospheric contamination is probably feasible in slowly rotating early-type stars just above $1.4$\,M$_\odot$. This result supports our finding that cooler early-type stars (also lower in mass) are better suited for studying the accretion of atoms and ions onto their photospheres. Due to their more intense stellar winds, hotter stars complicate the detection of chemical species such as oxygen and carbon in stellar spectra. Our model suggests that stars with effective temperatures in the range $6,000 - 8,000$\,K are optimal for such observations; temperatures above and below this range make detecting accretion signatures challenging.
\end{enumerate}


\section*{Acknowledgments}

The authors thank Sandipan Borthakur and Heleri Ramler for their insightful input and helpful discussions. We also extend our sincere thanks to the anonymous referee for their thorough review and constructive comments, which greatly helped to improve the clarity and reliability of this work. Their support has been greatly appreciated. This project has received funding from \fundingAgency{European Union's Horizon Europe research and innovation programme} under grant agreement no. \fundingNumber{101079231} (EXOHOST) and from the \fundingAgency{United Kingdom Research and Innovation (UKRI) Horizon Europe Guarantee Scheme} with grant No. \fundingNumber{10051045}. A.A. acknowledges support by the \fundingAgency{Estonian Research Council} grants \fundingNumber{PRG2159} and \fundingNumber{RVTT7}.


\subsection*{Conflict of interest}

The authors declare that they have no potential conflict of interest.






\bibliography{References.bib}


\newpage
\onecolumn
\appendix  

\section{Tables}\label{Appendix:tables}

\begin{longtable}{c c c c c c} 
\caption{Chosen parameters for calculation of $\beta$ ratios and velocity boosts. Age and surface gravity were takes as constants: Age = $20$ Myr and log g = 4.5 (because surface gravity does not affect SEDs strongly, so they were taken as constant).
The parameters marked with an asterisk ($^{*}$) corresponds to the parameters adopted for $\beta$ Pictoris, used for comparison with observed $\beta$ ratio values. For this star we assumed $M_\star$ = $1.75$\,M$_{\odot}$, $R\star = 1.53$\,R${_\odot}$, and $T{\mathrm{eff}} = 8,073$\,K, consistent with the values reported by \citet{DiFolco2004}}
\label{table:param}\\
\hline\hline
{\scriptsize \text{Mass [M}$_{\odot}$\text{]}} & {\scriptsize \text{Radius [R}$_{\odot}$\text{]}} & {\scriptsize \text{Effective Temperature [K]}}  \\
\hline 
\endfirsthead
\endhead
\hline \
\endlastfoot
$0.75$ & $0.576$ & $4000$   \\
$0.9$  & $0.655$ & $4500$   \\
$1.0$  & $0.720$ & $5000$   \\
$1.2$  & $0.933$ & $6000$   \\
$1.7$  & $1.03$ & $8000$   \\
$1.75^{*}$ & $1.53^{*}$ & $8000^{*}$ \\
$2.2$  & $1.15$ & $10,000$ \\
$3.2$  & $1.43$ & $13,000$ \\
$5.0$  & $1.97$ & $17,000$ \\
$6.2$  & $2.37$ & $19,000$ 
\end{longtable}


\newpage
\begin{landscape}
\small
\setlength{\tabcolsep}{5pt}
\fontsize{8}{10}
\begin{longtable}{p{0.2cm}*{9}{c}}             
\caption[An optional table caption ...]{$\beta$ ratios of all investigated atoms and ions across 9 different stellar effective temperatures. \label{table:beta}}\\
\hline\hline \\
{\scriptsize \text{Ion}} &
{\scriptsize \text{$4000$ K}} &
{\scriptsize \text{$4500$ K}} &
{\scriptsize \text{$5000$ K}} &
{\scriptsize \text{$6000$ K}} &
{\scriptsize \text{$8000$ K}} &
{\scriptsize \text{$10,000$ K}} &
{\scriptsize \text{$13,000$ K}} &
{\scriptsize \text{$17,000$ K}} &
{\scriptsize \text{$19,000$ K}} \\
\hline \\
\endfirsthead
\caption[]{(continued)}\\
\hline\hline \\
{\scriptsize \text{Ion}} &
{\scriptsize \text{$4000$ K}} &
{\scriptsize \text{$4500$ K}} &
{\scriptsize \text{$5000$ K}} &
{\scriptsize \text{$6000$ K}} &
{\scriptsize \text{$8000$ K}} &
{\scriptsize \text{$10,000$ K}} &
{\scriptsize \text{$13,000$ K}} &
{\scriptsize \text{$17,000$ K}} &
{\scriptsize \text{$19,000$ K}} \\
\hline \\
\endhead
\hline \\
{\scriptsize \text{'$0$' means that the value is lower than $10^{-30}$.}} \\ {\scriptsize \text{'-' means that there is no available data.}}\\ {\scriptsize \text{'$^f$' means that line data is only from forbidden transitions.}} \\
\endlastfoot
\ion{H}{I} &
  $(3.6 \pm 0.2) 10^{-13}$ &
  $(3.4 \pm 0.2)10^{-12}$ &
  $(3.9 \pm 0.2)10^{-10}$ &
  $(9.2 \pm 0.5)10^{-7}$ &
  $(2.2 \pm 0.1)10^{-3}$ &
  $(5.9 \pm 0.3)10^{-1}$ &
  $80.0 \pm 4.0$ &
  $3,750 \pm 187$ &
  $15,052 \pm 753$ \\
\ion{He}{I}&
 $0$&
 $0$&
 $0$&
  $(1.2 \pm 0.05)10^{-23}$ &
  $(1.8 \pm 0.09)10^{-16}$ &
  $(1.7 \pm 0.08)10^{-11}$ &
  $(5.6 \pm 0.3)10^{-7}$ &
  $(3.5 \pm 0.2)10^{-3}$ &
  $(5.3 \pm 0.3)10^{-2}$ \\
\ion{He}{II}&
 $0$&
 $0$&
 $0$&
 $0$&
 $0$&
  $(3.0 \pm 0.2)10^{-26}$ &
  $(6.3 \pm 0.2)10^{-19}$ &
  $(8.5 \pm 0.4)10^{-13}$ &
  $(8.1 \pm 0.4)10^{-11}$ \\
\ion{Li}{I}&
  $28.3 \pm 1.5$ &
  $69.0 \pm 3.6$ &
  $128 \pm 7$ &
  $382 \pm 20$ &
  $852 \pm 44$ &
  $1,367 \pm 70$ &
  $2,323 \pm 119$ &
  $4,316 \pm 222$ &
  $5,983 \pm 307$ \\
\ion{Li}{II}&
 $0$&
 $0$&
 $0$&
 $0$&
 $0$&
 $0$&
 $0$&
  $(1.2 \pm 0.06)10^{-22}$ &
  $(7.0 \pm 0.4)10^{-20}$ \\
\ion{Li}{III}&
 $0$&
 $0$&
 $0$&
 $0$&
 $0$&
 $0$&
 $0$&
 $0$&
 $0$\\
\ion{Be}{I}&
  $(1.4 \pm 0.2)10^{-4}$ &
  $(9.7 \pm 1.3)10^{-4}$ &
  $(4.1 \pm 0.6)10^{-2}$ &
  $3.3 \pm 0.5$ &
  $116 \pm 16$ &
  $560 \pm 77$ &
  $2,323 \pm 318$ &
  $8,706 \pm 119$ &
  $15,474 \pm 2120$ \\
\ion{Be}{II}&
  $(3.6 \pm 0.2)10^{-2}$ &
  $(3.3 \pm 0.2)10^{-1}$ &
  $2.6 \pm 0.1$ &
  $25.9 \pm 1.3$ &
  $126 \pm 6$ &
  $327 \pm 17$ &
  $1064 \pm 54$ &
  $3248 \pm 164$ &
  $5384 \pm 272$ \\
\ion{Be}{III}&
 $0$&
 $0$&
 $0$&
 $0$&
 $0$&
 $0$&
 $0$&
 $0$&
 $0$\\
\ion{B}{I}&
  $(2.7 \pm 0.2)10^{-5}$ &
  $(1.6 \pm 0.1)10^{-4}$ &
  $(3.6 \pm 0.3)10^{-3}$ &
  $(3.1 \pm 0.2)10^{-1}$ &
  $19.3 \pm 1.5$ &
  $232 \pm 17$ &
  $1,534 \pm 115$ &
  $6,757 \pm 507$ &
  $12,607 \pm 946$ \\
\ion{B}{II}&
  $(1.5 \pm 0.09)10^{-11}$ &
  $(4.9 \pm 0.3)10^{-11}$ &
  $(2.3 \pm 0.1)10^{-9}$ &
  $(5.6 \pm 0.3)10^{-6}$ &
  $(4.5 \pm 0.3)10^{-2}$ &
  $117 \pm 7$ &
  $1,134 \pm 70$ &
  $5,996 \pm 350$ &
  $11,153 \pm 685$ \\
\ion{B}{III}&
  $(1.7 \pm 0.8)10^{-6}$ &
  $(1.1 \pm 0.06)10^{-5}$ &
  $(7.8 \pm 0.4)10^{-4}$ &
  $(1.3 \pm 0.07)10^{-1}$ &
  $26.9 \pm 1.4$ &
  $135 \pm 7$ &  
  $585 \pm 30$ &
  $2,141 \pm 108$ &
  $3,743 \pm 189$ \\
\ion{C}{I}&
  $(9.2 \pm 0.6)10^{-10}$ &
  $(6.2 \pm 0.4)10^{-9}$ &
  $(5.3 \pm 0.4)10^{-7}$ &
  $(2.0 \pm 0.1)10^{-4}$ &
  $(4.6 \pm 0.3)10^{-1}$ &
  $53.6 \pm 3.9$ &
  $510 \pm 37$ &
  $2,801 \pm 203$ &
  $5,722 \pm 414$ \\
\ion{C}{II}&
  $(5.1 \pm 0.3)10^{-12}$ &
  $(3.2 \pm 0.2)10^{-11}$ &
  $(1.4 \pm 0.08)10^{-9}$ &
  $(5.9 \pm 0.4)10^{-7}$ &
  $(3.7 \pm 0.2)10^{-3}$ &
  $12.0 \pm 0.7$ &
  $126 \pm 8$ &
  $750 \pm 45$ &
  $1,668 \pm 100$ \\
\ion{C}{III}&
  $(1.4 \pm 0.1)10^{-17}$ &
  $(1.2 \pm 0.09)10^{-16}$ &
  $(1.9 \pm 0.1)10^{-14}$ &
  $(7.4 \pm 0.5)10^{-11}$ &
  $(1.5 \pm 0.1)10^{-6}$ &
  $(3.9 \pm 0.3)10^{-3}$ &
  $4.7 \pm 0.3$ &
  $351 \pm 25$ &
  $1,089 \pm 78$ \\
\ion{N}{I}&
  $(2.9 \pm 0.2)10^{-14}$ &
  $(1.3 \pm 0.09)10^{-13}$ &
  $(9.4 \pm 0.2610^{-12}$ &
  $(5.0 \pm 0.3)10^{-8}$ &
  $(1.8 \pm 0.1)10^{-4}$ &
  $(1.7 \pm 0.1)10^{-1}$ &
  $50.6 \pm 3.6$ &
  $758 \pm 53$ &
  $1,998 \pm 140$ \\
\ion{N}{II}&
  $(2.3 \pm 0.1)10^{-16}$ &
  $(1.4 \pm 0.07)10^{-15}$ &
  $(1.3 \pm 0.07)10^{-13}$ &
  $(2.9 \pm 0.2)10^{-10}$ &
  $(2.1 \pm 0.1)10^{-6}$ &
  $(3.3 \pm 0.2)10^{-3}$ &
  $14.9 \pm 0.8$ &
  $304 \pm 16$ &
  $741 \pm 39$ \\
\ion{N}{III}&
  $(4.3 \pm 0.3)10^{-18}$ &
  $(3.4 \pm 0.3)10^{-17}$ &
  $(4.6 \pm 0.3)10^{-15}$ &
  $(1.6 \pm 0.1)10^{-11}$ &
  $(2.8 \pm 0.2)10^{-7}$ &
  $(8.2 \pm 0.5)10^{-4}$ &
  $3.6 \pm 0.2$ &
  $181 \pm 11$ &
  $522 \pm 33$ \\
\ion{O}{I}&
  $(1.5 \pm 0.1)10^{-13}$ &
  $(4.4 \pm 0.3)10^{-13}$ &
  $(2.1 \pm 0.1)10^{-11}$ &
  $(8.2 \pm 0.5)10^{-8}$ &
  $(6.4 \pm 0.4)10^{-4}$ &
  $1.8 \pm 0.1$ &
  $25.9 \pm 1.7$ &
  $216 \pm 15$ &
  $524 \pm 35$ \\
\ion{O}{II}&
  $(4.5 \pm 0.3)10^{-26}$ &
  $(1.03 \pm 0.07)10^{-24}$ &
  $(7.4 \pm 0.5)10^{-22}$ &
  $(4.5 \pm 0.3)10^{-17}$ &
  $(7.2 \pm 0.5)10^{-12}$ &
  $(2.7 \pm 0.2)10^{-7}$ &
  $(2.4 \pm 0.2)10^{-5}$ &
  $(1.38 \pm 0.09)10^{-1}$ &
  $1.02 \pm 0.07$ \\
\ion{O}{III}&
  $(1.7 \pm 0.1)10^{-26}$ &
  $(3.9 \pm 0.3)10^{-25}$ &
  $(2.8 \pm 0.2)10^{-22}$ &
  $(1.7 \pm 0.1)10^{-17}$ &
  $(2.8 \pm 0.2)10^{-12}$ &
  $(1.1 \pm 0.08)10^{-7}$ &
  $(9.9 \pm 0.8)10^{-5}$ &
  $(6.1 \pm 0.5)10^{-2}$ &
  $(4.8 \pm 0.4)10^{-1}$ \\
\ion{F}{I}&
  $(5.2 \pm 0.4)10^{-19}$ &
  $(6.5 \pm 0.2)10^{-18}$ &
  $(4.2 \pm 0.3)10^{-16}$ &
  $(6.2 \pm 0.4)10^{-12}$ &
  $(2.6 \pm 0.2)10^{-7}$ &
  $(6.8 \pm 0.4)10^{-4}$ &
  $(2.0 \pm 0.1)10^{-1}$ &
  $29.7 \pm 2.0$ &
  $123 \pm 8$ \\
\ion{F}{II}&
  - &
  - &
  - &
  $(4.0 \pm 2.1)10^{-23}$ &
  $(3.4 \pm 1.9)10^{-16}$ &
  $(2.9 \pm 1.5)10^{-11}$ &
  $(5.9 \pm 3.1)10^{-7}$ &
  $(2.9 \pm 1.6)10^{-3}$ &
  $(4.1 \pm 2.2)10^{-2}$ \\
\ion{F}{III}&
  - &
  - &
  - &
  - &
  - &
  - &
  - &
  - &
  - \\
\ion{Ne}{I}&
 $0$ &
  $(3.2 \pm 0.2)10^{-28}$ &
  $(5.5 \pm 0.3)10^{-25}$ &
  $(1.3 \pm 0.08)10^{-19}$ &
  $(8.7 \pm 0.5)10^{-14}$ &
  $(4.1 \pm 0.3)10^{-9}$ &
  $(1.1 \pm 0.08)10^{-5}$ &
  $(1.4 \pm 0.08)10^{-2}$ &
  $(1.3 \pm 0.08)10^{-1}$ \\
\ion{Ne}{II}&
 $0$&
 $0$&
 $0$&
 $0$&
  $(8.8 \pm 0.1)10^{-23}$ &
  $(3.9 \pm 0.4)10^{-17}$ &
  $(2.7 \pm 0.3)10^{-12}$ &
  $(5.7 \pm 0.6)10^{-8}$ &
  $(1.2 \pm 0.1)10^{-6}$ \\
\ion{Ne}{III}&
 $0$&
 $0$&
 $0$&
  $(1.3 \pm 0.09)10^{-29}$&
  $(4.0 \pm 0.3)10^{-21}$ &
  $(1.2 \pm 0.09)10^{-15}$ &
  $(4.4 \pm 0.3)10^{-11}$ &
  $(5.6 \pm 0.4)10^{-7}$ &
  $(1.0 \pm 0.07)10^{-5}$ \\
\ion{Na}{I}&
  $4.6 \pm 0.3$ &
  $15.0 \pm 0.9$ &
  $36.6 \pm 2.2$ &
  $133 \pm 8$ &
  $351 \pm 21$ &
  $611 \pm 36$ &
  $1,064 \pm 63$ &
  $1,998 \pm 117$ &
  $2,775 \pm 163$ \\
\ion{Na}{II}&
 $0$&
 $0$&
 $0$&
 $0$&
  $(4.2 \pm 0.7)10^{-28}$ &
  $(1.1 \pm 0.2)10^{-21}$ &
  $(1.1 \pm 0.2)10^{-15}$ &
  $(1.7 \pm 0.3)10^{-10}$ &
  $(7.2 \pm 1.2)10^{-9}$ \\
\ion{Na}{III}&
 $0$&
 $0$&
 $0$&
 $0$&
  $(5.9 \pm 0.4)10^{-28}$ &
  $(1.2 \pm 0.09)10^{-21}$ &
  $(9.5 \pm 0.7)10^{-16}$ &
  $(1.3 \pm 0.09)10^{-10}$ &
  $(5.2 \pm 0.4)10^{-9}$ \\
\ion{Mg}{I}&
  $(4.0 \pm 0.5)10^{-3}$ &
  $(1.8 \pm 0.2)10^{-2}$ &
  $(1.4 \pm 0.2)10^{-1}$ &
  $5.6 \pm 0.7$ &
  $115 \pm 14$ &
  $385 \pm 47$ &
  $1,411 \pm 174$ &
  $4,669 \pm 574$ &
  $7,851 \pm 965$ \\
\ion{Mg}{II}&
  $(3.1 \pm 0.2)10^{-3}$ &
  $(1.6 \pm 0.1)10^{-2}$ &
  $(9.3 \pm 0.7)10^{-2}$ &
  $1.2 \pm 0.09$ &
  $25.6 \pm 1.9$ &
  $148 \pm 11$ &
  $624 \pm 45$ &
  $2,149 \pm 156$ &
  $3,616 \pm 262$ \\
\ion{Mg}{III}&
 $0$&
 $0$&
 $0$&
 $0$&
 $0$&
 $0$&
  $(6.3 \pm 1.1)10^{-27}$ &
  $(1.5 \pm 0.3)10^{-19}$ &
  $(5.9 \pm 1.0)10^{-17}$ \\
\ion{Al}{I}&
  $(4.4 \pm 0.3)10^{-2}$ &
  $(1.7 \pm 0.1)10^{-1}$ &
  $(8.2 \pm 0.6)10^{-1}$ &
  $7.7 \pm 0.6$ &
  $58.1 \pm 4.2$ &
  $183 \pm 13$ &
  $565 \pm 41$ &
  $1,727 \pm 125$ &
  $2,880 \pm 209$ \\
\ion{Al}{II}&
  $(1.8 \pm 0.9)10^{-8}$ &
  $(1.4 \pm 0.07)10^{-7}$ &
  $(6.1 \pm 0.3)10^{-6}$ &
  $(1.2 \pm 0.6)10^{-3}$ &
  $2.8 \pm 0.1$ &
  $156 \pm 8$ &
  $918 \pm 47$ &
  $4,227 \pm 220$ &
  $7,873 \pm 409$ \\
\ion{Al}{III}&
  $(9.3 \pm 0.5)10^{-8}$ &
  $(7.2 \pm 0.4)10^{-7}$ &
  $(7.9 \pm 0.4)10^{-5}$ &
  $(2.2 \pm 0.1)10^{-2}$ &
  $14.7 \pm 0.8$ &
  $109 \pm 6$ &
  $489 \pm 26$ &
  $1,827 \pm 99$ &
  $3,256 \pm 176$ \\
\ion{Si}{I}&
  $(3.5 \pm 0.5)10^{-5}$ &
  $(2.0 \pm 0.3)10^{-4}$ &
  $(3.7 \pm 0.6)10^{-3}$ &
  $(2.3 \pm 0.3)10^{-1}$ &
  $13.4 \pm 2.0$ &
  $94.8 \pm 14.5$ &
  $440 \pm 67$ &
  $1,696 \pm 259$ &
  $3,015 \pm 460$ \\
\ion{Si}{II}&
  $(4.3 \pm 0.5)10^{-10}$ &
  $(3.3 \pm 0.3)10^{-9}$ &
  $(2.1 \pm 0.2)10^{-7}$ &
  $(5.1 \pm 0.6)10^{-5}$ &
  $(5.9 \pm 0.6)10^{-2}$ &
  $17.9 \pm 1.9$ &
  $508 \pm 54$ &
  $4,082 \pm 436$ &
  $9,133 \pm 975$ \\
\ion{Si}{III}&
  $(6.6 \pm 1.4)10^{-14}$ &
  $(3.7 \pm 0.8)10^{-13}$ &
  $(3.0 \pm 0.6)10^{-11}$ &
  $(1.3 \pm 0.3)10^{-7}$ &
  $(4.2 \pm 0.9)10^{-4}$ &
  $(1.2 \pm 0.2)10^{-1}$ &
  $30.7 \pm 6.4$ &
  $877 \pm 181$ &
  $2,363 \pm 489$ \\
\ion{P}{I}&
  $(1.0 \pm 0.2)10^{-8}$ &
  $(8.3 \pm 0.1)10^{-8}$ &
  $(9.9 \pm 0.2)10^{-6}$ &
  $(3.4 \pm 0.6)10^{-3}$ &
  $4.4 \pm 0.7$ &
  $39.5 \pm 6.8$ &
  $186 \pm 31$ &
  $738 \pm 125$ &
  $1,330 \pm 225$ \\
\ion{P}{II}&
  $(3.5 \pm 0.9)10^{-12}$ &
  $(1.8 \pm 0.5)10^{-11}$ &
  $(5.2 \pm 1.5)10^{-10}$ &
  $(5.1 \pm 1.5)10^{-7}$ &
  $(1.0 \pm 0.3)10^{-3}$ &
  $1.2 \pm 0.3$ &
  $13.0 \pm 3.7$ &
  $66.3 \pm 18.8$ &
  $132 \pm 38$ \\
\ion{P}{III}&
  $(7.7 \pm 3.9)10^{-14}$ &
  $(2.5 \pm 1.3)10^{-13}$ &
  $(1.3 \pm 0.6)10^{-11}$ &
  $(3.9 \pm 2.0)10^{-8}$ &
  $(3.3 \pm 1.7)10^{-4}$ &
  $1.1 \pm 0.5$ &
  $11.0 \pm 5.5$ &
  $53.8 \pm 27.1$ &
  $107 \pm 54$ \\
\ion{S}{I}&
  $(4.6 \pm 0.4)10^{-9}$ &
  $(4.0 \pm 0.4)10^{-8}$ &
  $(4.1 \pm 0.4)10^{-6}$ &
  $(1.2 \pm 0.1)10^{-3}$ &
  $1.2 \pm 0.1$ &
  $32.9 \pm 3.0$ &
  $258 \pm 24$ &
  $1,410 \pm 130$ &
  $2,917 \pm 269$ \\
\ion{S}{II}&
  $(1.8 \pm 0.2)10^{-14}$ &
  $(5.4 \pm 0.7)10^{-14}$ &
  $(2.5 \pm 0.3)10^{-12}$ &
  $(1.3 \pm 0.2)10^{-8}$ &
  $(9.0 \pm 0.1)10^{-5}$ &
  $(1.1 \pm 0.2)10^{-1}$ &
  $9.5 \pm 1.2$ &
  $67.8 \pm 8.3$ &
  $143 \pm 18$ \\
\ion{S}{III}&
  $(1.3 \pm 0.08)10^{-15}$ &
  $(5.2 \pm 0.3)10^{-15}$ &
  $(3.2 \pm 0.2)10^{-13}$ &
  $(2.1 \pm 0.1)10^{-9}$ &
  $(8.5 \pm 0.5)10^{-6}$ &
  $(4.3 \pm 0.3)10^{-3}$ &
  $3.0 \pm 0.2$ &
  $60.3 \pm 3.8$ &
  $166 \pm 10$ \\
\ion{Cl}{I}&
  $(4.6 \pm 0.8)10^{-13}$ &
  $(1.5 \pm 0.2)10^{-12}$ &
  $(7.1 \pm 1.2)10^{-11}$ &
  $(2.0 \pm 0.3)10^{-7}$ &
  $(1.7 \pm 0.3)10^{-3}$ &
  $5.5 \pm 0.9$ &
  $61.5 \pm 10.2$ &
  $353 \pm 59$ &
  $724 \pm 120$ \\
\ion{Cl}{II}&
  $(7.3 \pm 1.3)10^{-18}$ &
  $(5.0 \pm 0.9)10^{-17}$ &
  $(5.3 \pm 1.0)10^{-15}$ &
  $(1.3 \pm 0.2)10^{-11}$ &
  $(1.1 \pm 0.2)10^{-7}$ &
  $(1.8 \pm 0.3)10^{-4}$ &
  $(8.6 \pm 1.6)10^{-1}$ &
  $20.1 \pm 3.7$ &
  $49.8 \pm 9.2$ \\
\ion{Cl}{III}&
  $(2.9 \pm 1.3)10^{-17}$ &
  $(2.5 \pm 1.1)10^{-16}$ &
  $(3.8 \pm 1.7)10^{-14}$ &
  $(1.3 \pm 0.6)10^{-10}$ &
  $(1.8 \pm 0.8)10^{-6}$ &
  $(4.6 \pm 2.0)10^{-3}$ &
  $12.5 \pm 5.5$ &
  $710 \pm 309$ &
  $2,154 \pm 939$ \\
\ion{Ar}{I}&
  $(5.0 \pm 0.4)10^{-17}$ &
  $(3.8 \pm 0.3)10^{-16}$ &
  $(4.5 \pm 0.4)10^{-14}$ &
  $(1.2 \pm 0.1)10^{-10}$ &
  $(1.1 \pm 0.1)10^{-6}$ &
  $(2.2 \pm 0.2)10^{-3}$ &
  $9.4 \pm 0.8$ &
  $298 \pm 26$ &
  $793 \pm 70$ \\
\ion{Ar}{II}&
  $(4.5 \pm 1.2)10^{-21}$ &
  $(3.6 \pm 1.0)10^{-20}$ &
  $(4.9 \pm 1.3)10^{-18}$ &
  $(2.2 \pm 0.6)10^{-14}$ &
  $(7.5 \pm 2.0)10^{-10}$ &
  $(1.4 \pm 0.4)10^{-6}$ &
  $(3.5 \pm 0.9)10^{-4}$ &
  $(3.2 \pm 0.8)10^{-2}$ &
  $(1.9 \pm 0.5)10^{-1}$ \\
\ion{Ar}{III}&
  $(5.2 \pm 1.0)10^{-26}$ &
  $(1.0 \pm 0.2)10^{-24}$ &
  $(5.1 \pm 0.9)10^{-22}$ &
  $(1.9 \pm 0.3)10^{-17}$ &
  $(1.8 \pm 0.3)10^{-12}$ &
  $(6.2 \pm 1.1)10^{-8}$ &
  $(3.8 \pm 0.7)10^{-5}$ &
  $(1.6 \pm 0.3)10^{-2}$ &
  $(1.1 \pm 0.2)10^{-1}$ \\
\ion{K}{I}&
  $10.0 \pm 0.6$ &
  $19.4 \pm 1.2$ &
  $33.9 \pm 2.1$ &
  $93.0 \pm 5.6$ &
  $185 \pm 11$ &
  $279 \pm 17$ &
  $458 \pm 28$ &
  $823 \pm 50$ &
  $1,130 \pm 68$ \\
\ion{K}{II}&
 $0$&
 $0$&
 $0$&
  $(1.1 \pm 0.4)10^{-23}$ &
  $(1.1 \pm 0.3)10^{-16}$ &
  $(9.0 \pm 2.9)10^{-12}$ &
  $(2.0 \pm 0.6)10^{-7}$ &
  $(1.0 \pm 0.3)10^{-3}$ &
  $(1.5 \pm 0.5)10^{-2}$ \\
\ion{K}{III}&
 $0$&
  $(1.6 \pm 0.4)10^{-28}$ &
  $(2.1 \pm 0.5)10^{-25}$ &
  $(3.3 \pm 0.8)10^{-20}$ &
  $(1.4 \pm 0.3)10^{-14}$ &
  $(5.9 \pm 1.5)10^{-10}$ &
  $(1.1 \pm 0.3)10^{-6}$ &
  $(1.0 \pm 0.3)10^{-3}$ &
  $(9.2 \pm 2.4)10^{-3}$ \\
\ion{Ca}{I}&
  $(7.2 \pm 0.9)10^{-1}$ &
  $3.4 \pm 0.4$ &
  $12.7 \pm 1.5$ &
  $75.7 \pm 9.2$ &
  $375 \pm 46$ &
  $893 \pm 109$ &
  $1,713 \pm 209$ &
  $3,498 \pm 426$ &
  $5,058 \pm 6.16$ \\
\ion{Ca}{II}&
  $(1.9 \pm 0.4)10^{-1}$ &
  $(6.7 \pm 1.2)10^{-1}$ &
  $2.5 \pm 0.5$ &
  $20.0 \pm 3.6$ &
  $141 \pm 26$ &
  $402 \pm 74$ &
  $858 \pm 158$ &
  $1,785 \pm 328$ &
  $2,582 \pm 474$ \\
\ion{Ca}{III}&
 $0$&
 $0$&
 $0$&
 $0$&
  $(2.4 \pm 0.2)10^{-24}$ &
  $(9.0 \pm 0.8)10^{-19}$ &
  $(9.5 \pm 0.8)10^{-14}$ &
  $(5.6 \pm 0.5)10^{-9}$ &
  $(1.8 \pm 0.2)10^{-7}$ \\
\ion{Sc}{I}&
  $(6.4 \pm 0.4)10^{-1}$ &
  $2.2 \pm 0.1$ &
  $6.7 \pm 0.4$ &
  $42.3 \pm 2.5$ &
  $213 \pm 13$ &
  $540 \pm 32$ &
  $1,213 \pm 73$ &
  $2,797 \pm 168$ &
  $4,224 \pm 253$ \\
\ion{Sc}{II}&
  $(1.2 \pm 0.1)10^{-1}$ &
  $(5.1 \pm 0.5)10^{-1}$ &
  $1.9 \pm 0.2$ &
  $14.5 \pm 1.3$ &
  $58.0 \pm 5.2$ &
  $143 \pm 13$ &
  $429 \pm 38$ &
  $1,217 \pm 108$ &
  $1,942 \pm 172$ \\
\ion{Sc}{III}&
  $(8.7 \pm 2.2)10^{-11}$ &
  $(6.2 \pm 1.5)10^{-10}$ &
  $(4.4 \pm 1.1)10^{-8}$ &
  $(2.3 \pm 0.6)10^{-5}$ &
  $(2.4 \pm 0.6)10^{-2}$ &
  $6.8 \pm 1.7$ &
  $36.1 \pm 9.0$ &
  $149 \pm 37$ &
  $268 \pm 67$ \\
\ion{Ti}{I}&
  $(2.1 \pm 0.1)10^{-1}$ &
  $(8.2 \pm 0.5)10^{-1}$ &
  $2.9 \pm 0.2$ &
  $20.9 \pm 1.3$ &
  $99.2 \pm 5.9$ &
  $253.7 \pm 15$ &
  $709 \pm 43$ &
  $1,947 \pm 117$ &
  $3,095 \pm 185$ \\
\ion{Ti}{II}&
  $(3.4 \pm 0.2)10^{-2}$ &
  $(1.8 \pm 0.1)10^{-1}$ &
  $(9.7 \pm 0.7)10^{-1}$ &
  $9.3 \pm 0.7$ &
  $46.1 \pm 3.3$ &
  $125 \pm 9$ &
  $414 \pm 29$ &
  $1,258 \pm 89$ &
  $2,052 \pm 145$ \\
\ion{Ti}{III}&
  $(4.3 \pm 2.2)10^{-14}$ &
  $(1.3 \pm 0.7)10^{-13}$ &
  $(6.4 \pm 3.2)10^{-12}$ &
  $(2.8 \pm 1.4)10^{-8}$ &
  $(2.5 \pm 1.2)10^{-4}$ &
  $(9.8 \pm 4.9)10^{-1}$ &
  $13.9 \pm 7.0$ &
  $69.1 \pm 34.7$ &
  $134 \pm 67$ \\
\ion{V}{I}&
  $(3.8 \pm 0.2)10^{-1}$ &
  $1.4 \pm 0.08$ &
  $4.8 \pm 0.3$ &
  $38.2 \pm 2.1$ &
  $190 \pm 11$ &
  $484 \pm 27$ &
  $1,406 \pm 78$ &
  $3.901 \pm 217$ &
  $6,204 \pm 346$ \\
\ion{V}{II}&
  $(4.2 \pm 0.3)10^{-2}$ &
  $(3.3 \pm 0.3)10^{-1}$ &
  $7.9 \pm 0.6$ &
  $167 \pm 13$ &
  $1,640 \pm 131$ &
  $5,807 \pm 465$ &
  $22,245 \pm 1,780$ &
  $75,691 \pm 6,055$ &
  $128,552 \pm 10,284$ \\
\ion{V}{III}&
  - &
  - &
  - &
  - &
  - &
  - &
  - &
  - &
  - \\
\ion{Cr}{I}&
  $(2.3 \pm 0.2)10^{-1}$ &
  $(8.3 \pm 0.6)10^{-1}$ &
  $2.8 \pm 0.2$ &
  $20.6 \pm 1.4$ &
  $90.6 \pm 6.1$ &
  $213 \pm 14$ &
  $517 \pm 35$ &
  $1,274 \pm 86$ &
  $1,946 \pm 131$ \\
\ion{Cr}{II}&
  $(2.2 \pm 0.6)10^{-7}$ &
  $(1.6 \pm 0.5)10^{-6}$ &
  $(1.0 \pm 0.3)10^{-4}$ &
  $(1.8 \pm 0.5)10^{-2}$ &
  $3.6 \pm 1.0$ &
  $18.0 \pm 5.1$ &
  $77.5 \pm 22.0$ &
  $281 \pm 80 $ &
  $492 \pm 140$ \\
\ion{Cr}{III}&
  - &
  - &
  - &
  - &
  - &
  - &
  - &
  - &
  - \\
\ion{Mn}{I}&
  $(6.2 \pm 0.5)10^{-2}$ &
  $(2.2 \pm 0.2)10^{-1}$ &
  $(6.7 \pm 0.6)10^{-1}$ &
  $4.3 \pm 0.4$ &
  $34.3 \pm 2.9$ &
  $138 \pm 12$ &
  $475 \pm 40$ &
  $1,507 \pm 128$ &
  $2,491 \pm 212$ \\
\ion{Mn}{II}&
  $(2.0 \pm 0.2)10^{-4}$ &
  $(1.4 \pm 0.1)10^{-3}$ &
  $(3.4 \pm 0.3)10^{-2}$ &
  $(9.7 \pm 0.7)10^{-1}$ &
  $16.2 \pm 1.2$ &
  $66.6 \pm 5.0$ &
  $271 \pm 20$ &
  $1,086 \pm 82$ &
  $2,027 \pm 152$ \\
\ion{Mn}{III}&
  - &
  - &
  - &
  - &
  - &
  - &
  - &
  - &
  - \\
\ion{Fe}{I}&
  $(1.3 \pm 0.08)10^{-2}$ &
  $(5.1 \pm 0.3)10^{-2}$ &
  $(2.6 \pm 0.2)10^{-1}$ &
  $3.6 \pm 0.2$ &
  $34.9 \pm 2.1$ &
  $123 \pm 7.3$ &
  $462 \pm 28$ &
  $1,571 \pm 94$ &
  $2,682 \pm 160$ \\
\ion{Fe}{II}&
  $(8.7 \pm 0.08)10^{-5}$ &
  $(6.1 \pm 0.5)10^{-4}$ &
  $(1.5 \pm 0.1)10^{-2}$ &
  $(4.6 \pm 0.4)10^{-1}$ &
  $11.8 \pm 1.1$ &
  $57.7 \pm 5.2$ &
  $259 \pm 23$ &
  $1,101 \pm 99$ &
  $2,058 \pm 184$ \\
\ion{Fe}{III}&
  $(1.4 \pm 0.2)10^{-16}$ &
  $(6.6 \pm 1.0)10^{-16}$ &
  $(5.1 \pm 0.8)10^{-14}$ &
  $(5.5 \pm 0.9)10^{-10}$ &
  $(3.4 \pm 0.5)10^{-6}$ &
  $(1.1 \pm 0.2)10^{-2}$ &
  $4.9 \pm 0.8$ &
  $40.2 \pm 6.3$ &
  $88.6 \pm 13.9$ \\
\ion{Co}{I}&
  $(5.3 \pm 0.5)10^{-3}$ &
  $(2.4 \pm 0.2)10^{-2}$ &
  $(1.1 \pm 0.1)10^{-1}$ &
  $1.4 \pm 0.1$ &
  $17.8 \pm 1.6$ &
  $71.7 \pm 6.5$ &
  $281 \pm 26$ &
  $1,010 \pm 92$ &
  $1,763 \pm 161$ \\
\ion{Co}{II}&
  $(4.5 \pm 0.5)10^{-8}$ &
  $(3.3 \pm 0.3)10^{-7}$ &
  $(2.3 \pm 0.2)10^{-5}$ &
  $(4.5 \pm 0.5)10^{-3}$ &
  $1.1 \pm 0.1$ &
  $11.4 \pm 1.2$ &
  $63.8 \pm 6.6$ &
  $262 \pm 27$ &
  $473 \pm 49$ \\
\ion{Co}{III}&
  $(1.7 \pm 0.2)10^{-19}$ &
  $(1.2 \pm 0.1)10^{-18}$ &
  $(1.6 \pm 0.1)10^{-16}$ &
  $(7.1 \pm 0.7)10^{-13}$ &
  $(2.1 \pm 0.2)10^{-8}$ &
  $(6.2 \pm 0.6)10^{-5}$ &
  $(1.9 \pm 0.2)10^{-2}$ &
  $2.4 \pm 0.2$ &
  $13.3 \pm 1.2$ \\
\ion{Ni}{I}&
  $(2.7 \pm 0.2)10^{-3}$ &
  $(1.3 \pm 0.1)10^{-2}$ &
  $(7.0 \pm 0.6)10^{-2}$ &
  $1.2 \pm 0.1$ &
  $18.3 \pm 1.5$ &
  $72.0 \pm 5.8$ &
  $289 \pm 23$ &
  $1,049 \pm 85$ &
  $1,842 \pm 149$ \\
\ion{Ni}{II}&
  $(3.1 \pm 0.8)10^{-10}$ &
  $(2.6 \pm 0.7)10^{-9}$ &
  $(3.6 \pm 0.9)10^{-7}$ &
  $(1.3 \pm 0.3)10^{-4}$ &
  $(2.1 \pm 0.5)10^{-1}$ &
  $1.8 \pm 0.4$ &
  $8.2 \pm 2.1$ &
  $32.4 \pm 8.3$ &
  $58.8 \pm 15.0$ 
\end{longtable}
\end{landscape}

\begin{landscape}
\small
\setlength{\tabcolsep}{5pt}
\fontsize{8}{0}
\begin{longtable}{p{0.2cm}*{9}{c}} 
\caption{Ionisation Rates of Neutral Atoms at a distance $100$\,AU.}
\label{table:ion rate}\\
\hline\hline \\
{\scriptsize \text{Ion}} & 
{\scriptsize  $4000$\, {\scriptsize \text{K}}} &
{\scriptsize  $4500$\, {\scriptsize \text{K}}} &
{\scriptsize  $5000$\, {\scriptsize \text{K}}} &
{\scriptsize $6000$\, {\scriptsize \text{K}}} &
{\scriptsize $8000$\, {\scriptsize \text{K}}} &
{\scriptsize $10,000$\, {\scriptsize \text{K}}} &
{\scriptsize $13,000$\, {\scriptsize \text{K}}} &
{\scriptsize $17,000$\, {\scriptsize \text{K}}} &
{\scriptsize $19,000$\, {\scriptsize \text{K}}} \\
\hline \\ 
\endfirsthead
\hline \\
{\scriptsize \text{In units of [$10^{-9}$ /s]}} \\
\endlastfoot
$\ion{H}{I}$  &      &      &      &       &       & $0.468$ &$1.247$  &$4.563$ &$11.09$\\
$\ion{Li}{I}$ &$0.019$&$0.039$&$0.071$&$0.218$ &$0.680$ &$1.424$ &$3.796$ &$18.83$ &$58.25$ \\
$\ion{Be}{I}$ &      &      &      & $0.294$ &$0.917$  &$1.929$  &$5.131$ &$36.90$ &$135.6$ \\
$\ion{B}{I}$  &      &      &      &       &$1.166$  &$2.476$  &$6.561$ &$73.59$ &$304.4$ \\
$\ion{C}{I}$  &      &      &      &       &       &$3.086$  &$8.132$ &$141.3$  &$625.9$  \\
$\ion{N}{I}$  &      &      &      &       &       &       &$9.935$ &$255.4$  &$1,175$  \\
$\ion{O}{I}$  &      &      &      &       &       &$4.658$ &$12.07$ &$433.3$  &$2,038$  \\
$\ion{F}{I}$  &      &      &      &       &       &        &        &$693.9$ &$3,307$ \\
$\ion{Na}{I}$ & $0.211$ &$0.502$&$1.143$&$4.268$  &$16.11$  &$41.67$  &$140.5$  &$11,518$ &$56,149$ \\
$\ion{Mg}{I}$ &      &      &     &  $5.863$ &$22.39$  &$57.94$ &$198.3$ &$16,044$  &$78,206$  \\
$\ion{Al}{I}$ &      &      &$0.933$&$3.348$ &$12.22$  &$34.97$  &$95.34$  &$11,779$  &$57,678$  \\
$\ion{Si}{I}$ &      &      &      &       &$15.81$&$45.56$ &$125.5$ &$15,555$ &$76,181$  \\
$\ion{P}{I}$  &      &      &      &       &$20.26$  &$58.53$ &$162.9$ &$20,050$ &$98,196$ \\
$\ion{S}{I}$  &      &      &      &       &$25.69$  &$74.21$ &$208.7$ &$25,335$  &$124,066$  \\
$\ion{Cl}{I}$ &      &      &      &       &       &$92.91$  &$264.0$ &$31,396$ &$153,720$ \\
$\ion{Ar}{I}$ &      &      &      &       &       &       &$297.5$  &$35,377$  &$173,209$  \\
$\ion{K}{I}$  &$0.604$&$1.473$&$3.414$&$12.91$ &$49.25$ &$140.6$ &$407.5$ &$46,200$ &$226,085$ \\
$\ion{Ca}{I}$ &$0.721$&$1.771$&$4.129$ &$15.69$ &$60.03$ &$170.4$ &$498.4$ &$55,049$ &$269,310$ \\
$\ion{Sc}{I}$ &$0.855$&$2.115$&$4.956$&$18.89$ &$72.49$&$204.6$&$603.2$&$64,773$&$316,769$\\
$\ion{Ti}{I}$ &      &$2.507$ &$5.900$ &$22.55$ &$86.74$ &$243.1$ &$723.3$&$75,070$&$366,961$\\
$\ion{V}{I}$  &      &$1.213$&$2.865$&$10.97$&$42.29$&$115.5$&$353.0$&$29,978$&$146,098$\\
$\ion{Cr}{I}$ &      &$3.462$ &$8.201$ &$31.48$ &$121.5$ &$323.9$ &$1,015$ &$70,501$ &$342,305$ \\
$\ion{Mn}{I}$ &      &      & $9.579$ &$36.83$  &$142.3$  &$347.8$ &$1,189$ &$28,369$ &$132,335$ \\
$\ion{Fe}{I}$ &      &      &      &$42.65$  &$164.9$ &$390.7$ &$1,378$ &$26,638$ &$122,444$ \\
$\ion{Co}{I}$ &      &      &      &$49.43$  &$191.3$  &$442.3$ &$1,599$ &$25,368$ &$114,619$ \\
$\ion{Ni}{I}$ &      &      &      &         &$3,924$  &$8,111$  &$34,984$ &$113,671$ &$332,117$ 
\end{longtable}
\end{landscape}


\begin{landscape}
\small
\setlength{\tabcolsep}{5pt}
\fontsize{8}{0}
\begin{longtable}{p{0.2cm}*{9}{c}} 
\caption{Velocity boosts of Neutral Atoms before ionisation}
\label{table:velocity}\\
\hline\hline \\
{\scriptsize \text{Ion}} & 
{\scriptsize  $4000$\, {\scriptsize \text{K}}} &
{\scriptsize  $4500$\, {\scriptsize \text{K}}} &
{\scriptsize  $5000$\, {\scriptsize \text{K}}} &
{\scriptsize $6000$\, {\scriptsize \text{K}}} &
{\scriptsize $8000$\, {\scriptsize \text{K}}} &
{\scriptsize $10,000$\, {\scriptsize \text{K}}} &
{\scriptsize $13,000$\, {\scriptsize \text{K}}} &
{\scriptsize $17,000$\, {\scriptsize \text{K}}} &
{\scriptsize $19,000$\, {\scriptsize \text{K}}} \\
\hline \\ 
\endfirsthead
\hline \\
{\scriptsize \text{In units of [km/s]}} \\
\endlastfoot
$\ion{H}{I}$  &      &      &      &       &       & $1.65$ &$122$  &$2,437$ &$4,992$\\
$\ion{Li}{I}$ &$643$&$954$&$1,074$&$1,245$ &$1,262$ &$1,252$ &$1,161$ &$680$ &$378$ \\
$\ion{Be}{I}$ &      &      &      & $8.11$ &$127$  &$379$  &$859$ &$700$ &$420$ \\
$\ion{B}{I}$  &      &      &      &       &$16.8$  &$122$  &$444$ &$272$ &$152$ \\
$\ion{C}{I}$  &      &      &      &       &       &$22.7$  &$119$ &$58.8$  &$33.6$  \\
$\ion{N}{I}$  &      &      &      &       &       &       &$9.67$ &$8.80$  &$6.25$  \\
$\ion{O}{I}$  &      &      &      &       &       &$0.49$ &$4.06$ &$1.48$  &$0.94$  \\
$\ion{F}{I}$  &      &      &      &       &       &       &       &$0.13$ &$0.14$ \\
$\ion{Na}{I}$ & $9.73$ &$15.9$&$19.0$&$22.1$  &$22.0$  &$19.1$  &$14.4$  &$0.51$ &$0.18$ \\
$\ion{Mg}{I}$ &      &      &     &  $0.68$ &$5.17$  &$8.67$ &$13.5$ &$0.86$  &$0.37$  \\
$\ion{Al}{I}$ &      &      &$0.52$&$1.63$ &$4.80$  &$6.83$  &$11.3$  &$0.43$  &$0.18$  \\
$\ion{Si}{I}$ &      &      &      &       &$0.85$&$2.72$ &$6.65$ &$0.32$ &$0.15$  \\
$\ion{P}{I}$  &      &      &      &       &$0.22$  &$0.88$ &$2.17$ &$0.11$ &$0.05$ \\
$\ion{S}{I}$  &      &      &      &       &$0.05$  &$0.58$ &$2.34$ &$0.17$  &$0.09$  \\
$\ion{Cl}{I}$ &      &      &      &       &       &$0.08$  &$0.44$ &$0.03$ &$0.02$ \\
$\ion{Ar}{I}$ &      &      &      &       &       &       &$0.06$  &$0.02$  &$0.02$  \\
$\ion{K}{I}$  &$7.34$&$7.05$&$5.89$&$5.13$ &$3.78$ &$2.59$ &$2.13$ &$0.05$ &$0.02$ \\
$\ion{Ca}{I}$ &$0.45$&$1.04$&$1.82$ &$3.43$ &$6.29$ &$6.84$ &$6.52$ &$0.19$ &$0.07$ \\
$\ion{Sc}{I}$ &$0.33$&$0.56$&$0.80$&$1.59$ &$2.30$&$3.44$&$3.82$&$0.13$&$0.05$\\
$\ion{Ti}{I}$ &      &$0.17$ &$0.29$ &$0.66$ &$1.15$ &$1.36$ &$1.86$&$0.08$&$0.03$\\
$\ion{V}{I}$  &      &$0.61$&$0.99$&$2.48$&$4.53$&$5.47$&$7.56$&$0.39$&$0.16$\\
$\ion{Cr}{I}$ &      &$0.13$ &$0.20$ &$0.47$ &$0.75$ &$0.86$ &$0.97$ &$0.05$ &$0.02$ \\
$\ion{Mn}{I}$ &      &      & $0.04$ &$0.08$  &$0.24$  &$0.52$ &$0.76$ &$0.16$ &$0.07$ \\
$\ion{Fe}{I}$ &      &      &      &$0.06$  &$0.21$ &$0.41$ &$0.64$ &$0.17$ &$0.08$ \\
$\ion{Co}{I}$ &      &      &      &$0.02$  &$0.09$  &$0.21$ &$0.33$ &$0.12$ &$0.06$ \\
$\ion{Ni}{I}$ &      &      &      &   &$0.01$  &$0.01$  &$0.02$ &$0.03$ &$0.02$ 
\end{longtable}
\end{landscape}

\begin{landscape}
\small
\setlength{\tabcolsep}{5pt}
\fontsize{8}{0}
\begin{longtable}{p{0.5cm} c c c p{0.5cm} c c c}
\caption{$\beta$ ratio values from this work with photospheric emission and modified stellar EUV compared to Fernandez et al. 2006 results.}
\label{tab:comp fernandez}\\
\hline\hline \\
 {\scriptsize \text{Ion}} & {\scriptsize \text{Photospheric emission}} & {\scriptsize \text{Modified Stellar EUV}} & {\scriptsize \text{Fernandez}} & {\scriptsize \text{Ion}} & {\scriptsize \text{Photospheric emission}} & {\scriptsize \text{Modified Stellar EUV}} & {\scriptsize \text{Fernandez}} \\
\hline \\
\endfirsthead
\hline \\
{\scriptsize \text{'$0$' means that the value is lower than $10^{-30}$.}} \\ {\scriptsize \text{'-' means that there is no available linelist data.}} \\
\endlastfoot
$\ion{H}{I}$ &
  $(2.5 \pm 0.1)10^{-3}$ &
  $4.8 \pm 0.2$ &
  \multicolumn{1}{l|}{$(1.6 \pm 0.1)10^{-3}$} &
  $\ion{S}{I}$ &
  $1.4 \pm 0.1$ &
  $1.4 \pm 0.1$ &
  $(5.6 \pm 0.9)10^{-1}$ \\
$\ion{He}{I}$ &
  $(2.1 \pm 0.1)10^{-16}$ &
  $(9.2 \pm 0.5)10^{-5}$ &
  \multicolumn{1}{l|}{$0$} &
  $\ion{S}{II}$ &
  $(1.0 \pm 0.1)10^{-4}$ &
  $(2.7 \pm 0.3)10^{-4}$ &
  $(9.0 \pm 1.0)10^{-5}$ \\
$\ion{He}{II}$ &
  $0$ &
  $(5.6 \pm 0.03)10^{-4}$ &
  \multicolumn{1}{l|}{-} &
  $\ion{S}{III}$ &
  $(9.6 \pm 0.07)10^{-6}$ &
  $(3.0 \pm 0.2)10^{-5}$ &
  $(2.0 \pm 1.0)10^{-4}$ \\
$\ion{Li}{I}$ &
  $967 \pm 50$ &
  $967 \pm 50$ &
  \multicolumn{1}{l|}{$900 \pm 40$} &
  $\ion{Cl}{I}$ &
  $(1.9 \pm 0.3)10^{-3}$ &
  $(1.9 \pm 0.3)10^{-3}$ &
  $(2.3 \pm 0.4)10^{-3}$ \\
$\ion{Li}{II}$ &
  $0$ &
  $(4.4 \pm 0.2)10^{-5}$ &
  \multicolumn{1}{l|}{$0$} &
  $\ion{Cl}{II}$ &
  $(1.2 \pm 0.2)10^{-7}$ &
  $(7.7 \pm 1.4)10^{-7}$ &
  $(3.7 \pm 0.4)10^{-7}$ \\
$\ion{Li}{III}$ &
  $0$ &
  $(1.8 \pm 0.1)10^{-6}$ &
  \multicolumn{1}{l|}{-} &
  $\ion{Cl}{III}$ &
  $(2.1 \pm 0.9)10^{-6}$ &
  $(3.5 \pm 0.2)10^{-5}$ &
  $(3.0 \pm 2.0)10^{-6}$ \\
$\ion{Be}{I}$ &
  $131 \pm 18$ &
  $131 \pm 18$ &
  \multicolumn{1}{l|}{$62 \pm 7$} &
  $\ion{Ar}{I}$ &
  $(1.3 \pm 0.1)10^{-6}$ &
  $(3.2 \pm 0.3)10^{-5}$ &
  $(1.7 \pm 0.3)10^{-6}$ \\
$\ion{Be}{II}$ &
  $143 \pm 7$ &
  $144 \pm 7$ &
  \multicolumn{1}{l|}{$124 \pm 6$} &
  $\ion{Ar}{II}$ &
  $(8.6 \pm 2.3)10^{-10}$ &
  $(1.9 \pm 0.5)10^{-6}$ &
  $0$ \\
$\ion{Be}{III}$ &
  $0$ &
  $(4.3 \pm 0.2)10^{-6}$ &
  \multicolumn{1}{l|}{$0$} &
  $\ion{Ar}{III}$ &
  $(2.0 \pm 0.4)10^{-12}$ &
  $(3.8 \pm 0.7)10^{-6}$ &
  $(1.5 \pm 0.2)10^{-7}$ \\
$\ion{B}{I}$ &
  $22.0 \pm 1.7$ &
  $22.1 \pm 1.7$ &
  \multicolumn{1}{l|}{$30 \pm 10$} &
  $\ion{K}{I}$ &
  $210 \pm 13$ &
  $210 \pm 13$ &
  $200 \pm 20$ \\
$\ion{B}{II}$ &
  $(5.2 \pm 0.3)10^{-2}$ &
  $(5.0 \pm 0.3)10^{-2}$ &
  \multicolumn{1}{l|}{$(7 \pm 4)10^{-2}$} &
  $\ion{K}{II}$ &
  $(1.2 \pm 0.4)10^{-16}$ &
  $(2.1 \pm 0.7)10^{-4}$ &
  - \\
$\ion{B}{III}$ &
  $30.5 \pm 1.5$ &
  $30.5 \pm 1.5$ &
  \multicolumn{1}{l|}{$19 \pm 1$} &
  $\ion{K}{III}$ &
  $(1.5 \pm 0.4)10^{-14}$ &
  $(1.5 \pm 0.4)10^{-5}$ &
  $(4.4 \pm 0.2)10^{-4}$ \\
$\ion{C}{I}$ &
  $(5.2 \pm 0.4)10^{-1}$ &
  $(5.2 \pm 0.4)10^{-1}$ &
  \multicolumn{1}{l|}{$(3.3 \pm 0.1)10^{-2}$} &
  $\ion{Ca}{I}$ &
  $425 \pm 52$ &
  $426 \pm 52$ &
  $330 \pm 40$ \\
$\ion{C}{II}$ &
  $(4.2 \pm 0.3)10^{-3}$ &
  $(4.6 \pm 0.3)10^{-3}$ &
  \multicolumn{1}{l|}{$(2.3 \pm 0.2)10^{-3}$} &
  $\ion{Ca}{II}$ &
  $160 \pm 29$ &
  $166 \pm 31$ &
  $50.0 \pm 10.0$ \\
$\ion{C}{III}$ &
  $(1.7 \pm 0.1)10^{-6}$ &
  $(6.6 \pm 0.5)10^{-4}$ &
  \multicolumn{1}{l|}{$(8.5 \pm 0.9)10^{-3}$} &
  $\ion{Ca}{III}$ &
  $(2.8 \pm 0.2)10^{-24}$ &
  $(1.3 \pm 0.1)10^{-4}$ &
  - \\
$\ion{N}{I}$ &
  $(2.1 \pm 0.1)10^{-4}$ &
  $(2.1 \pm 0.2)10^{-4}$ &
  \multicolumn{1}{l|}{$(2.1 \pm 0.1)10^{-4}$} &
  $\ion{Sc}{I}$ &
  $242 \pm 15$ &
  $239 \pm 14$ &
  $220 \pm 20$ \\
$\ion{N}{II}$ &
  $(2.4 \pm 0.1)10^{-6}$ &
  $(7.2 \pm 0.4)10^{-5}$ &
  \multicolumn{1}{l|}{$(7.5 \pm 0.5)10^{-6}$} &
  $\ion{Sc}{II}$ &
  $65.9 \pm 5.8$ &
  $66.1 \pm 5.9$ &
  $1300 \pm 400$ \\
$\ion{N}{III}$ &
  $(3.2 \pm 0.2)10^{-7}$ &
  $(5.9 \pm 0.4)10^{-5}$ &
  \multicolumn{1}{l|}{$(7.0 \pm 1.0)10^{-4}$} &
  $\ion{Sc}{III}$ &
  $(2.8 \pm 0.7)10^{-2}$ &
  $(2.8 \pm 0.7)10^{-2}$ &
  $(9.0 \pm 3.0)10^{-2}$ \\
$\ion{O}{I}$ &
  $(7.3 \pm 0.5)10^{-4}$ &
  $(7.8 \pm 0.5)10^{-4}$ &
  \multicolumn{1}{l|}{$(3.3 \pm 0.2)10^{-4}$} &
  $\ion{Ti}{I}$ &
  $113 \pm 7$ &
  $112 \pm 7$ &
  $97.0 \pm 5$ \\
$\ion{O}{II}$ &
  $(8.2 \pm 0.6)10^{-12}$ &
  $(5.4 \pm 0.4)10^{-5}$ &
  \multicolumn{1}{l|}{$(3.1 \pm 0.7)10^{-9}$} &
  $\ion{Ti}{II}$ &
  $52.3 \pm 3.7$ &
  $52.4 \pm 3.7$ &
  $28 \pm 2$ \\
$\ion{O}{III}$ &
  $(3.2 \pm 0.2)10^{-12}$ &
  $(2.7 \pm 0.2)10^{-4}$ &
  \multicolumn{1}{l|}{$(6.5 \pm 0.6)10^{-7}$} &
  $\ion{Ti}{III}$ &
  $(2.8 \pm 0.1)10^{-4}$ &
  $(2.8 \pm 0.1)10^{-4}$ &
  $(5.0 \pm 0.1)10^{-4}$ \\
$\ion{F}{I}$ &
  $(7.7 \pm 0.5)10^{-8}$ &
  $(5.9 \pm 0.4)10^{-6}$ &
  \multicolumn{1}{l|}{$0$} &
  $\ion{V}{I}$ &
  $216 \pm 12$ &
  $216 \pm 12$ &
  $72 \pm 4$ \\
$\ion{F}{II}$ &
  $(4.0 \pm 2.2)10^{-16}$ &
  $(3.2 \pm 1.7)10^{-5}$ &
  \multicolumn{1}{l|}{$(3.5 \pm 0.9)10^{-6}$} &
  $\ion{V}{II}$ &
  $1,862 \pm 149$ &
  $1,862 \pm 149$ &
  $4.4 \pm 0.2$ \\
$\ion{F}{III}$ &
  - &
  - &
  \multicolumn{1}{l|}{$(5.0 \pm 1.0)10^{-9}$} &
  $\ion{V}{III}$ &
  - &
  - &
  $0$ \\
$\ion{Ne}{I}$ &
  $(9.8 \pm 0.6)10^{-14}$ &
  $(6.4 \pm 0.4)10^{-6}$ &
  \multicolumn{1}{l|}{$0$} &
  $\ion{Cr}{I}$ &
  $103 \pm 7$ &
  $103 \pm 7$ &
  $93 \pm 5$ \\
$\ion{Ne}{II}$ &
  $(1.0 \pm 0.1)10^{-22}$ &
  $(1.1 \pm 0.1)10^{-6}$ &
  \multicolumn{1}{l|}{$0$} &
  $\ion{Cr}{II}$ &
  $4.0 \pm 1.2$ &
  $4.0 \pm 1.1$ &
  $(6.0 \pm 3.0)10^{-7}$ \\
$\ion{Ne}{III}$ &
  $(4.6 \pm 0.3)10^{-21}$ &
  $(2.7 \pm 0.2)10^{-5}$ &
  \multicolumn{1}{l|}{$(9.0 \pm 2.0)10^{-8}$} &
  $\ion{Cr}{III}$ &
  - &
  - &
  - \\
$\ion{Na}{I}$ &
  $399 \pm 23$ &
  $400 \pm 24$ &
  \multicolumn{1}{l|}{$360 \pm 20$} &
  $\ion{Mn}{I}$ &
  $38.9 \pm 3.3$ &
  $38.9 \pm 2.4$ &
  $28 \pm 3$ \\
$\ion{Na}{II}$ &
  $(4.8 \pm 0.8)10^{-28}$ &
  $(4.0 \pm 0.6)10^{-5}$ &
  \multicolumn{1}{l|}{$0$} &
  $\ion{Mn}{II}$ &
  $18.4 \pm 1.4$ &
  $18.4 \pm 1.4$ &
  $7 \pm 1$ \\
$\ion{Na}{III}$ &
  $(6.7 \pm 0.5)10^{-28}$ &
  $(1.2 \pm 0.8)10^{-5}$ &
  \multicolumn{1}{l|}{$0$} &
  $\ion{Mn}{III}$ &
  - &
  - &
  - \\
$\ion{Mg}{I}$ &
  $130 \pm 16$ &
  $130 \pm 16$ &
  \multicolumn{1}{l|}{$74 \pm 8$} &
  $\ion{Fe}{I}$ &
  $39.6 \pm 2.4$ &
  $39.5 \pm 2.4$ &
  $27 \pm 2$ \\
$\ion{Mg}{II}$ &
  $29.1 \pm 2.1$ &
  $29.1 \pm 2.1$ &
  \multicolumn{1}{l|}{$9 \pm 24$} &
  $\ion{Fe}{II}$ &
  $13.4 \pm 1.2$ &
  $13.4 \pm 1.2$ &
  $5.0 \pm 0.3$ \\
$\ion{Mg}{III}$ &
  $0$ &
  $(1.2 \pm 0.2)10^{-5}$ &
  \multicolumn{1}{l|}{$0$} &
  $\ion{Fe}{III}$ &
  $(3.9 \pm 0.6)10^{-6}$ &
  $(1.4 \pm 0.2)10^{-5}$ &
  $(3.0 \pm 0.6)10^{-7}$ \\
$\ion{Al}{I}$ &
  $66.0 \pm 4.8$ &
  $66.0 \pm 4.8$ &
  \multicolumn{1}{l|}{$53 \pm 6$} &
  $\ion{Co}{I}$ &
  $20.3 \pm 1.8$ &
  $20.2 \pm 1.8$ &
  $16 \pm 1$ \\
$\ion{Al}{II}$ &
  $3.2 \pm 0.2$ &
  $5.5 \pm 0.3$ &
  \multicolumn{1}{l|}{$(3.6 \pm 0.5)10^{-1}$} &
  $\ion{Co}{II}$ &
  $1.2 \pm 0.1$ &
  $1.2 \pm 0.1$ &
  $0$ \\
$\ion{Al}{III}$ &
  $16.7 \pm 0.9$ &
  $16.7 \pm 0.9$ &
  \multicolumn{1}{l|}{$12 \pm 1$} &
  $\ion{Co}{III}$ &
  $(2.4 \pm 0.2)10^{-8}$ &
  $(1.1 \pm 0.1)10^{-5}$ &
  $(4.0 \pm 2)10^{-7}$ \\
$\ion{Si}{I}$ &
  $15.2 \pm 2.3$ &
  $15.4 \pm 2.3$ &
  \multicolumn{1}{l|}{$6.0 \pm 0.6$} &
  $\ion{Ni}{I}$ &
  $20.8 \pm 1.8$ &
  $20.1 \pm 1.7$ &
  $26 \pm 2$ \\
$\ion{Si}{II}$ &
  $(6.7 \pm 0.7)10^{-2}$ &
  $(6.7 \pm 0.7)10^{-2}$ &
  \multicolumn{1}{l|}{$9 \pm 9$} &
  $\ion{Ni}{II}$ &
  $(2.3 \pm 0.6)10^{-1}$ &
  $(2.4 \pm 0.6)10^{-1}$ &
  $(7.0 \pm 2.0)10^{-2}$ \\
$\ion{Si}{III}$ &
  $(4.7 \pm 1.0)10^{-4}$ &
  $(4.7 \pm 1.0)10^{-4}$ &
  \multicolumn{1}{l|}{$(5.8 \pm 0.6)10^{-4}$} &
  $\ion{Ni}{III}$ &
  - &
  - &
  $(3.0 \pm 2.0)10^{-7}$ \\
$\ion{P}{I}$ &
  $5.0 \pm 0.8$ &
  $5.0 \pm 0.8$ &
  \multicolumn{1}{l|}{$3.4 \pm 0.6$} &
   &
   &
   &
   \\
$\ion{P}{II}$ &
  $(1.2 \pm 0.3)10^{-3}$ &
  $(1.2 \pm 0.3)10^{-3}$ &
  \multicolumn{1}{l|}{$(2.2 \pm 0.3)10^{-3}$} &
   &
   &
   &
   \\
$\ion{P}{III}$ &
  $(3.8 \pm 1.9)10^{-4}$ &
  $(3.8 \pm 1.9)10^{-4}$ &
  \multicolumn{1}{l|}{$(5.0 \pm 2.0)10^{-4}$} &
   &
   &
   & 
\end{longtable}
\end{landscape}


\begin{longtable}{p{0.2cm}*{3}{c}} 
\caption{Velocity boost values from this work with photospheric emission and modified stellar EUV compared to Fernandez et al. 2006 results.}
\label{tab:comp Fernandez velocity}\\
\hline\hline
{\scriptsize \text{Ion}} & {\scriptsize \text{Photospheric emission}} & {\scriptsize \text{Modified Stellar EUV}} & {\scriptsize \text{Fernandez}}           \\
\hline 
\endfirsthead
\hline
{\scriptsize \text{In units of [km/s]}} \\
\endlastfoot
$\ion{Li}{I}$ &$1,262$                 &$1,250$                &$0.11$               \\
$\ion{Be}{I}$ &$127$                  &$126$                 &$58$                 \\
$\ion{B}{I}$  &$16.8$                   &$16.6$                  &$0.41$               \\
$\ion{Na}{I}$ &$22.0$                  &$20.5$                 &$3.3$                \\
$\ion{Mg}{I}$ &$5.2$                   &$4.8$                  &$1.1$                \\
$\ion{Al}{I}$ &$4.8$                   &$4.5$                  & $5.0 \cdot 10^{-4}$ \\
$\ion{Si}{I}$ &$0.85$                  &$0.80$                 &$0.02$               \\
$\ion{P}{I}$  &$0.22$                  &$0.21$                 &$8.7$                \\
$\ion{S}{I}$  &$0.05$                  &$0.04$                 &$1.5$                \\
$\ion{K}{I}$  &$3.8$                   &$3.5$                  &$0.5$                \\
$\ion{Ca}{I}$ &$6.3$                   &$5.9$                  &$0.03$               \\
$\ion{Sc}{I}$ &$3.0$                  &$2.8$                 &$3.5$                \\
$\ion{Ti}{I}$ &$1.2$                   &$1.1$                  &$0.6$                \\
$\ion{V}{I}$  &$4.5$                  &$4.3$                 &$0.2$                \\
$\ion{Cr}{I}$ &$0.8$                   &$0.7$                  &$0.9$                \\
$\ion{Mn}{I}$ &$0.2$                  &$0.23$                 &$0.4$                \\
$\ion{Fe}{I}$ &$0.2$                   &$0.2$                  &$0.5$                \\
$\ion{Co}{I}$ &$0.09$                  &$0.09$                 &$0.8$                \\
$\ion{Ni}{I}$ &$0.005$                  &$0.004$                 &$0.3$               
\end{longtable}

\section{Figures}\label{appendix:figures}

\begin{figure}[p]
\centerline{\includegraphics[width=0.98\textwidth]{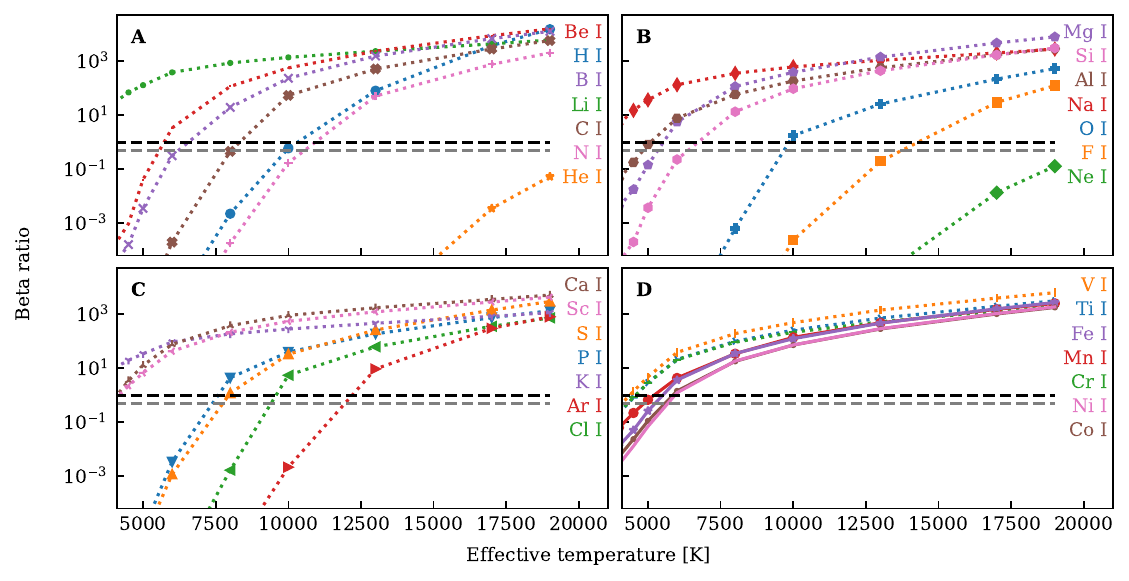}}
    \caption{$\beta$ ratios of the selected neutral atoms as a function of stellar effective temperature. The atoms are arranged across the panels in descending order of their \(\beta\) values at T$_{\rm eff}$ = $19,000$ \,K. The black dashed line ($\beta = 1$) indicates the equilibrium condition, where the radiative and gravitational forces acting on the atoms are balanced and grey dashed line ($\beta = 0.5$) indicates the $\beta$ value above which unhindered gas, when released from circular orbit, starts to migrate outwards.}
    \label{fig:allneut}
\end{figure}

\begin{figure} 
\centerline{\includegraphics[width=0.98\textwidth]{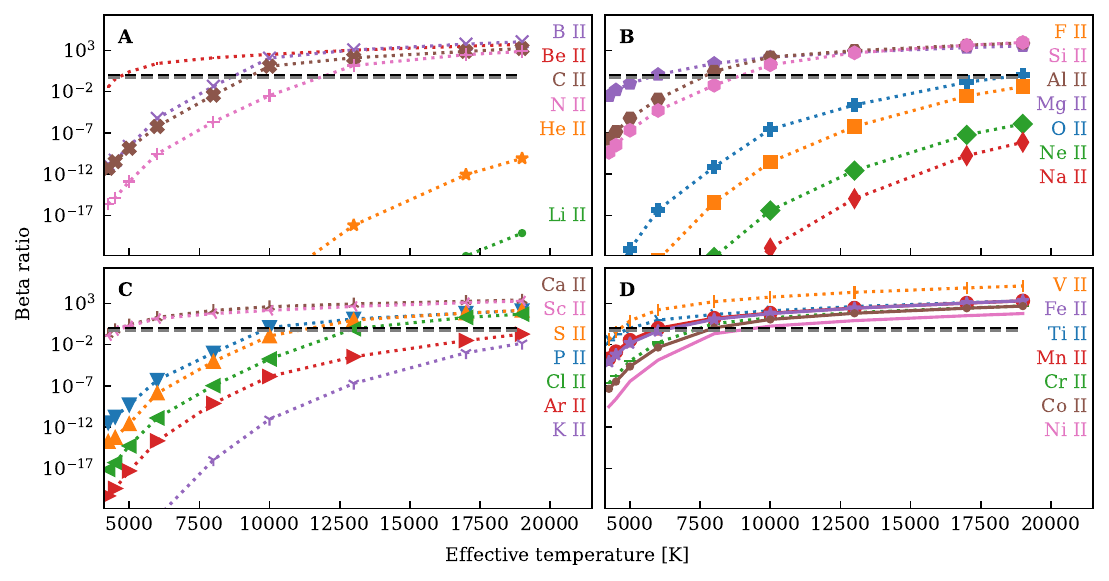}}
        \caption{Same as Figure \ref{fig:neut}, but for singly ionised species.}
        \label{fig:allsinglyionised}
\end{figure}
\begin{figure}
    \centerline{\includegraphics[width=0.98\textwidth]{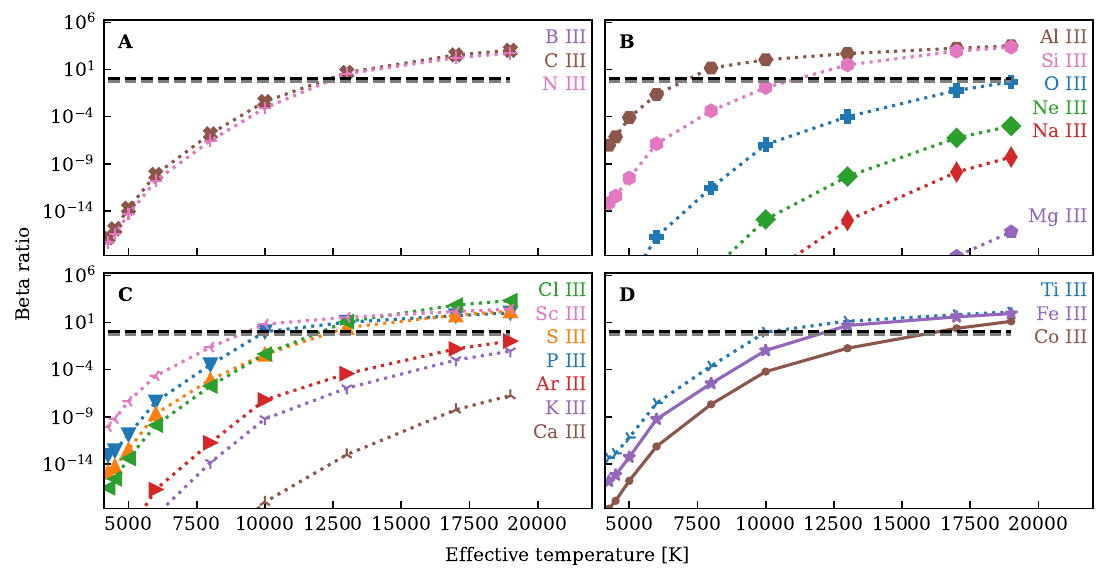}}
        \caption{Same as Figure \ref{fig:neut}, but for doubly ionised species.}
        \label{fig:alldoublyionised}
\end{figure}

\begin{figure}
    \centerline{\includegraphics[width=0.98\textwidth]{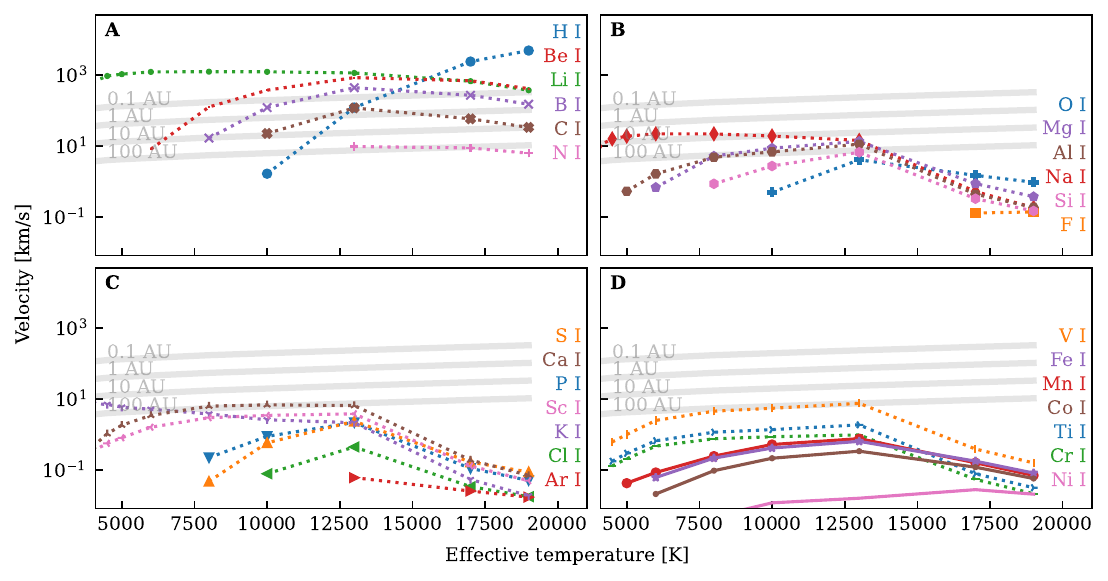}}
        \caption{Velocity boosts of $28$ neutral atoms with \(\beta > 0.5\) as a function of stellar effective temperature. The atoms are arranged across the panels in descending order of their \(\beta\) values at \(T_{\mathrm{eff}} = 19,000~\mathrm{K}\). Shown in grey are the escape velocities at distances of \(0.1\), \(1\), \(10\), and \(100~\mathrm{AU}\).}
        \label{fig:allvel}
\end{figure}

\end{document}